\documentclass{aa}  

\usepackage{graphicx}
\usepackage{txfonts}
\usepackage{lscape}
\usepackage{placeins}
\usepackage{booktabs}

\begin{document}

   \title{Probing faint radio activity from cores to filaments in two $z=2.24$ protoclusters with uGMRT }

   \author{Yuheng Zhang\inst{1,2} \email{yuheng@nju.edu.cn}
	\and Xianzhong Zheng\inst{3} \email{xzzheng@sjtu.edu.cn}
	\and Fangxia An\inst{4} \email{anfangxia@ynao.ac.cn}
	\and Tao Wang\inst{1,2} \email{taowang@nju.edu.cn}
	\and Qinghua Tan\inst{5,6} \email{qhtan@pmo.ac.cn}
    \and Dongdong Shi\inst{7} \email{ddshi@aust.edu.cn}
    \and Hao Peng\inst{5,6} \email{haopeng@pmo.ac.cn}
        }

\institute{
	School of Astronomy and Space Science, Nanjing University, Nanjing 210093, China
    \and Key Laboratory of Modern Astronomy and Astrophysics (Nanjing University), Ministry of Education, Nanjing 210093, China
    \and State Key Laboratory of Dark Matter Physics, Tsung-Dao Lee Institute, Shanghai Jiao Tong University, Shanghai, 201210, China
    \and Yunnan Observatories, Chinese Academy of Sciences, Kunming 650216, China
    \and Purple Mountain Observatory, Chinese Academy of Sciences, 10 Yuanhua Road, Nanjing, 210023, China 
    \and School of Astronomy and Space Science, University of Science and Technology of China, Hefei, Anhui 230026, China
    \and Center for Fundamental Physics, School of Mechanics and Optoelectronic Physics, Anhui University of Science and Technology, Huainan, Anhui 232001, China
}

   \date{Received September 30, 20XX}

  \abstract
   { We present deep 1.26\,GHz radio continuum observations of the two $z=2.24$ protoclusters BOSS1244 and BOSS1542 with the upgraded Giant Metrewave Radio Telescope (uGMRT) to investigate the radio emission of their member galaxies and its possible environmental dependence. Among 187 and 105 H$\alpha$ emitter (HAE) members in BOSS1244 and BOSS1542, we identify ten and two robust radio counterparts, respectively, together with five potential counterparts from forced photometry. Two HAEs and the massive quiescent galaxy QG1 in BOSS1244 show strong radio excesses relative to their H$\alpha$-based star formation rates (SFRs) and are identified as strong radio-excess AGN candidates. Median stacking of the remaining 173 and 102 radio-undetected HAE members yields positive signals at $2.8\sigma$ and $3.2\sigma$ in BOSS1244 and BOSS1542, respectively. The stacked radio luminosities are consistent with those expected from the median H$\alpha$-based SFRs, indicating that star formation can account for the average faint radio emission. After excluding the robust radio counterparts, we further divided the HAE members by the projected distance from QG1 in BOSS1244 and from the filament spine in BOSS1542. The largest stacked radio luminosities are found toward the denser regions in both systems, but the differences are not statistically significant, and stellar mass shows no significant dependence on environmental distance. These results identify strong radio-excess AGN candidates among the individually detected members, while the faint HAE population is on average consistent with radio emission from star formation. The similar environmental ordering in the two protoclusters is suggestive but remains tentative. Deeper radio observations at multiple frequencies and higher angular resolution are required to determine whether this apparent trend reflects weak AGN activity or variations in radio emission associated with star formation.
}
   
   \titlerunning{Faint radio activity in $z=2.24$ protoclusters}
   \maketitle
   \nolinenumbers

\section{Introduction}

Understanding how active galactic nuclei (AGNs) regulate the growth of massive galaxies remains a central challenge in galaxy formation. AGN feedback, particularly in its radio mode, is thought to play a key role in heating the surrounding gas and suppressing star formation, thereby shaping the high-mass end of the galaxy population \citep{Fabian2012, Heckman2014}. While luminous AGNs have been extensively studied, low-level nuclear activity is more difficult to identify and to connect directly to galaxy evolution. In particular, instantaneous AGN luminosity does not necessarily trace the cumulative impact of black hole growth and feedback \citep{Terrazas2020, Bluck2023}. Identifying faint AGN activity and separating it from emission powered by star formation are therefore important for tracing black hole growth in the general galaxy population.

Protoclusters at $z\sim2$--3 provide a unique environment to probe these processes during the peak epoch of galaxy and black hole growth. At these redshifts, galaxies reside in dense, rapidly assembling structures characterized by abundant cold gas supply, frequent interactions, and strong environmental variations from filamentary regions to forming cluster cores \citep{Overzier2016, Chiang2017}. Observational studies have shown that both star formation and AGN activity may be enhanced in the early assembly of protoclusters compared to the field, although with significant diversity across systems and environments \citep{Casey2016, Lehmer2009, Umehata2019}. Deep X-ray observations of the Spiderweb protocluster, for example, reveal widespread nuclear activity among its member galaxies \citep{Tozzi2022a}, while \citet{Shimakawa2024a} find that approximately half of its massive HAEs host X-ray AGNs and identify a population of AGN hosts with low star formation rates (SFRs). However, existing constraints are still dominated by individually identified AGNs and a limited number of well-studied protoclusters. The prevalence of faint radio activity among more typical protocluster galaxies, its physical origin, and its possible dependence on the internal environment remain poorly constrained.

A key unresolved question is how galaxy and nuclear activity vary across different internal environments of protoclusters, and how this variation connects to the early stages of cluster assembly. Theoretical models predict that dense environments, such as protocluster cores and filaments, can enhance AGN triggering through increased merger rates, gas inflows, and angular momentum loss \citep{HopkinsP2006,Angles-Alcazar2017}. Observationally, elevated AGN fractions have been reported in several high-redshift protoclusters \citep{Tozzi2022a,Vito2024,Chapman2026,IslaLlave2026}. However, the reported level of enhancement varies substantially among systems, ranging from strong AGN excesses to AGN fractions consistent with those in the field \citep{Digby-North2010,Gatica2024}. Radio continuum observations provide a complementary view of this activity because they are little affected by dust obscuration and trace synchrotron emission associated with both star formation and AGNs. This dual origin, however, also complicates the interpretation of faint radio emission. Radio excess relative to that expected from star formation, together with radio spectral and morphological information, can provide evidence for an AGN contribution \citep{Delvecchio2017,Delhaize2017,CalistroRivera2017}. A growing number of studies have applied radio observations to high-redshift protoclusters, revealing radio overdensities and AGN activity in a handful of systems \citep[e.g.,][]{Ao2017,Daddi2017,Chapman2026}. However,  systematic radio studies of the more numerous, individually faint member galaxies remain scarce, particularly those that probe how their radio emission changes between different environments within the same protocluster.

The protoclusters BOSS1244 and BOSS1542 at $z=2.24$ are among the most massive systems at cosmic noon. The significant overdensities of H$\alpha$ emitters (HAEs) and submillimeter galaxies (SMGs) indicate active star formation and mass assembly in these two fields \citep{ZhengX2021,ShiD2021,ZhangY2022}. 
Their internal structures, however, differ markedly: BOSS1244 contains several prominent density peaks and a central concentration hosting a pair of massive quiescent galaxies (QGs) \citep{ShiD2021,ShiD2024}, whereas BOSS1542 is dominated by an extended filamentary structure \citep{ZhengX2021,ShiD2021}.
The merger rates of massive HAEs are elevated in both systems relative to the field, with a higher value measured in BOSS1542 \citep{LiuS2023}. An excess of early-type galaxies around the central region of BOSS1244 further suggests that an emerging cluster core may already be developing there \citep{Golden-Marx2026}. These contrasting properties make them an ideal pair for investigating how environment shapes galaxy evolution and AGN activity within protoclusters. 
In this work, we present deep 1.26 GHz radio continuum observations of BOSS1244 and BOSS1542 obtained with the upgraded Giant Metrewave Radio Telescope \citep[uGMRT; ][]{Gupta2017}, and use them to characterize the radio emission of their HAE members. Combining individual radio diagnostics with median stacking, we further examine whether the faint radio emission varies with the internal protocluster environment.

This paper is organized as follows. In Sect.~\ref{sec:data}, we describe the uGMRT observations and data reduction. In Sect.~\ref{sec:results}, we introduce the HAE member samples and present their radio counterparts, radio diagnostics, and stacking analyses, including the dependence of the stacked emission on the internal protocluster environment. In Sect.~\ref{sec:discuss}, we discuss the origin of the radio emission and its possible environmental modulation, and we summarize our findings in Sect.~\ref{sec:summary}.
Throughout this work, we adopt a \citet{Chabrier2003} initial mass function (IMF) and assume a flat $\Lambda$CDM cosmology with $H_0 = 70\ \mathrm{km\ s^{-1}\ Mpc^{-1}}$ and $\Omega_\Lambda = 0.7$ \citep{Spergel2003,Spergel2007}.

\section{Data reduction and analysis}
\label{sec:data}

\begin{figure*}
    \begin{minipage}[c]{12.5cm}
    \hbox to 12.5cm{\includegraphics[width=6.15cm]{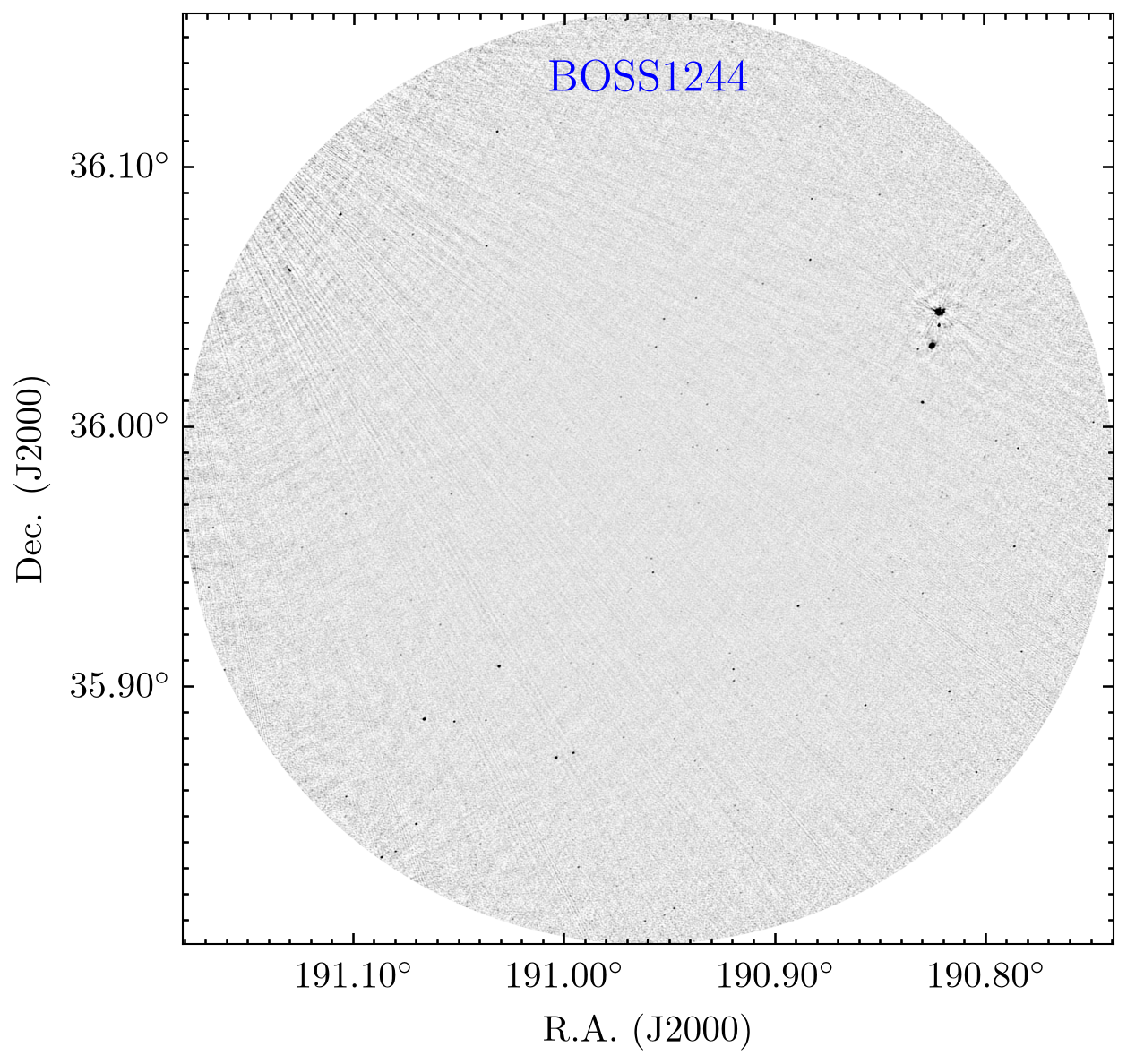}\hfil\includegraphics[width=6.15cm]{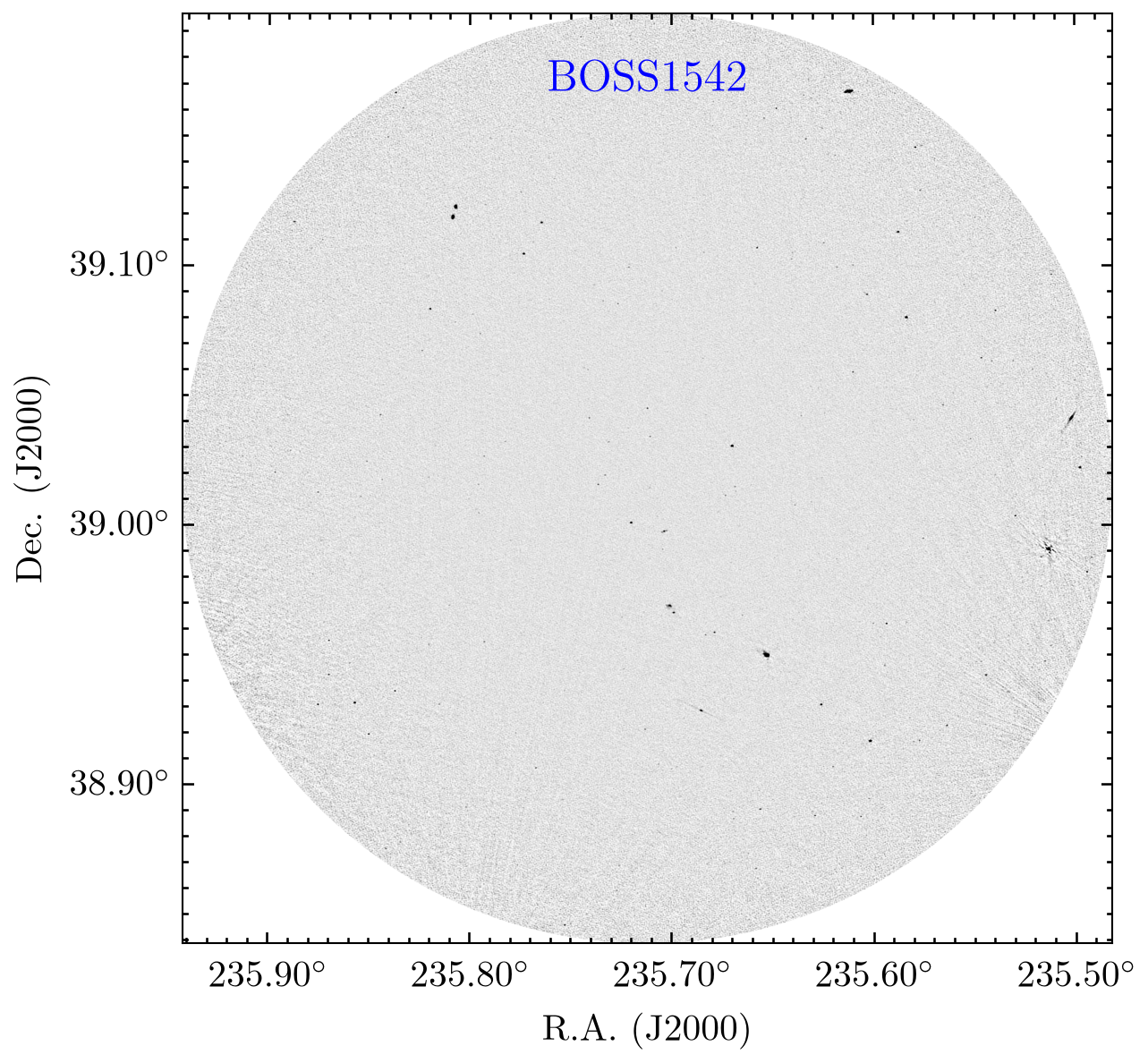}}
    \end{minipage}\hfill
    \begin{minipage}[c]{\dimexpr\textwidth-12.5cm-12pt\relax}
    \caption{Deep uGMRT images at 1.26\,GHz of BOSS1244 (left) and BOSS1542 (right) after primary-beam correction. The effective survey area is defined by pixels with a primary-beam response greater than 0.4, within which source extraction and all subsequent analyses were performed.}
    \label{fig:images}
    \end{minipage}
\end{figure*}

\subsection{uGMRT observations}

The uGMRT Band~5 observations of BOSS1244 and BOSS1542 were carried out on 2022 April 23 and 2022 December 9, respectively (project codes: 42\_054 and 43\_031; PI: Y.~Zhang). Both observations covered a nominal bandwidth of 400\,MHz from 1060 to 1460\,MHz, centered at 1260\,MHz. The observations alternated 35-min target scans with 8-min phase-calibrator scans, using 1227$+$365 for BOSS1244 and 1602$+$334 for BOSS1542. The flux-density and bandpass calibrators were 3C\,147 and 3C\,286 for BOSS1244, and 3C\,48 and 3C\,286 for BOSS1542. A total of 12 and 10 target scans were obtained for BOSS1244 and BOSS1542, respectively, with total observing time of approximately 11 and 9\,hr.

\subsection{Calibration and imaging}

The raw visibilities were converted into measurement sets and reduced with the Common Astronomy Software Applications package \citep[CASA v6.6.4.34;][]{CASA2022}, following standard calibration procedures\footnote{https://gmrt-tutorials.readthedocs.io/en/latest/index.html}.
Initial flagging removed shadowed data, zero-valued samples, quack intervals, short baselines, and the outer 3\% of channels at both band edges, together with manual flagging of evident radio frequency interference (RFI) and malfunctioning antennas. Additional automated RFI mitigation was performed using the CASA task \texttt{tfcrop} prior to calibration.

The absolute flux density scale was set using primary flux calibrators and transferred to the phase calibrator. Delay, bandpass, and complex gain solutions were derived and applied to all calibrators. We subsequently applied \texttt{rflag} to further suppress residual RFI, followed by a second round of calibration to ensure stable solutions. The final calibration tables were applied to the science target data, which were then split from the full dataset. A final \texttt{rflag} pass, supplemented by manual flagging, was then applied to the science target data. Owing to severe RFI contamination, approximately 46.2\% and 27.1\% of the science visibilities remained unflagged and were retained for subsequent imaging for BOSS1244 and BOSS1542, respectively.

We performed six rounds of phase-only self-calibration using the CAPTURE pipeline \citep[CAsa Pipeline-cum-Toolkit for Upgraded Giant Metrewave Radio Telescope data REduction;][]{Kale2021} to mitigate artifacts introduced by bright sources within the field of view. The band was divided into eight subbands, with the solution interval progressively shortened from 8, 4, 2, and 1\,min to 30\,s and one integration. Despite these direction-independent corrections, prominent radial artifacts persisted around the brightest sources.

We therefore applied the Python-based, CASA-compatible peeling framework of \citet{PengH2026} to the self-calibrated visibilities for direction-dependent calibration. The framework derives source-specific gain solutions for problematic bright sources while preserving the emission elsewhere in the field. Two directions were peeled in BOSS1244 and one in BOSS1542.

The final peeled visibilities were imaged using the CASA task \texttt{tclean}. We adopted the \texttt{wproject} gridder to account for wide-field noncoplanar baseline effects, with the number of $w$-planes determined automatically. Wideband deconvolution was performed using multiterm multifrequency synthesis  with two Taylor terms to model spectral variation across the observing band, together with Briggs weighting with a robust parameter of zero \citep{Briggs1995}. Deconvolution was carried out in two stages. A conservative mask was first generated with \texttt{auto-multithresh} at approximately $5\sigma$ and inspected manually; the resulting mask was then fixed and cleaning continued within it to approximately $3\sigma$. In BOSS1244, a spurious mask region associated with residual structure around the northeastern peeled source was excluded before the second stage.

The final restored beams are $2\farcs43\times2\farcs07$ at a position angle of $-66.6\degr$ for BOSS1244 and $2\farcs03\times1\farcs62$ at $89.8\degr$ for BOSS1542. The corresponding central off-source root-mean-square (rms) noise levels are 10.2 and 11.5\,$\mu$Jy\,beam$^{-1}$, respectively. We restricted all subsequent analysis to regions where the primary-beam response exceeds 0.4, corresponding to effective areas of 362.5 and 362.9\,arcmin$^2$. The final science images after primary beam correction are shown in Fig.~\ref{fig:images}.

We assessed the external flux-density scale and tested for any radial dependence in the recovered flux densities using compact, isolated sources with high signal-to-noise ratios (S/N). Radio counterparts were required to be mutual nearest neighbors within $3\arcsec$, with uGMRT and external catalog peak S/N of at least 10 and 5, respectively. Cross-matching with FIRST \citep{Becker1995} yielded only one suitable source in BOSS1244 and two in BOSS1542, which is insufficient for a statistical assessment. We therefore used 25 and 18 matches from LoTSS-DR3 \citep{Shimwell2026}. Their integrated flux densities were scaled from 144\,MHz to 1.26\,GHz assuming a spectral index of $\alpha=-0.57$ from broadband radio fits spanning approximately 150\,MHz to 1.25\,GHz \citep{SinhaSpectral2023}. The median ratios between the integrated flux densities measured with uGMRT and scaled LoTSS are $1.04_{-0.42}^{+0.86}$ and $0.90_{-0.22}^{+0.81}$ for BOSS1244 and BOSS1542, respectively. The broad scatter is expected in part from intrinsic variations in source spectral indices over this large frequency interval. To test for radial dependence independently of the absolute normalization in each field, we normalized the flux-density ratios by their field medians and fitted their logarithms as a function of angular distance from the phase center. 
The joint fit gives a slope of $k=0.03\pm0.03$ per arcmin with a p-value of 0.38, consistent with no measurable radial trend. Separate fits to the two fields lead to the same conclusion. We therefore find no evidence for a systematic radial bias in the measured flux densities within the adopted analysis region.

We additionally assessed the astrometric accuracy against the deep CFHT/WIRCam $K_{\rm s}$-band catalogs \citep{ZhengX2021}, matching high S/N and isolated radio sources within $1\arcsec$. This comparison revealed a systematic offset in BOSS1542 of  $-0.20\arcsec$ in R.A. and $-0.35\arcsec$ in Dec., defined as uGMRT minus $K_{\rm s}$. We therefore corrected the astrometry of the BOSS1542 image accordingly.  After correction, the measured offsets for the 40 BOSS1244 matches are $-0.05\pm0.15\arcsec$ in R.A. and $-0.13\pm0.15\arcsec$ in Dec., respectively. For the 32 BOSS1542 matches, the corresponding offsets are $0.00\pm0.14\arcsec$ in R.A. and $0.00\pm0.16\arcsec$ in Dec., respectively.

\section{Results}
\label{sec:results}

\begin{figure*}[t]
    \centering
    \begin{minipage}[t]{0.495\textwidth}
        \includegraphics[width=\linewidth]{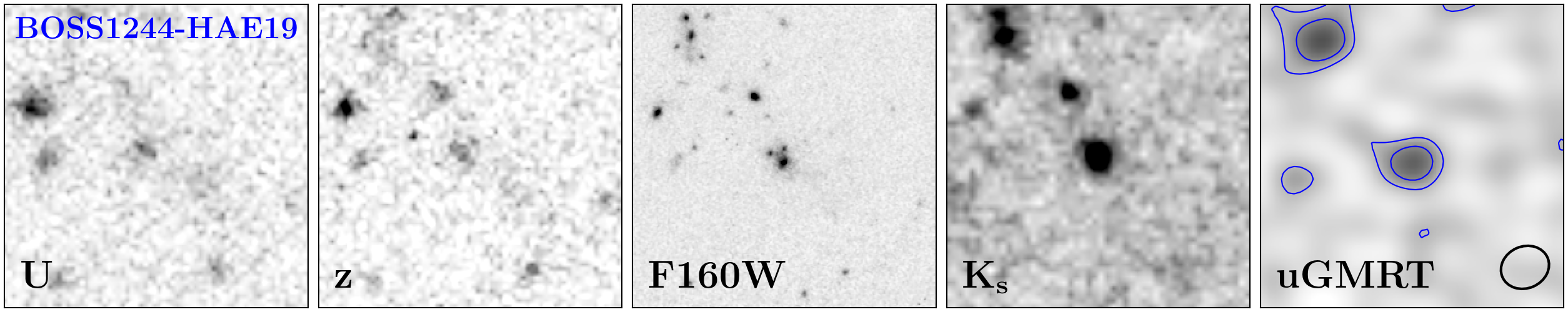}\par
        \includegraphics[width=\linewidth]{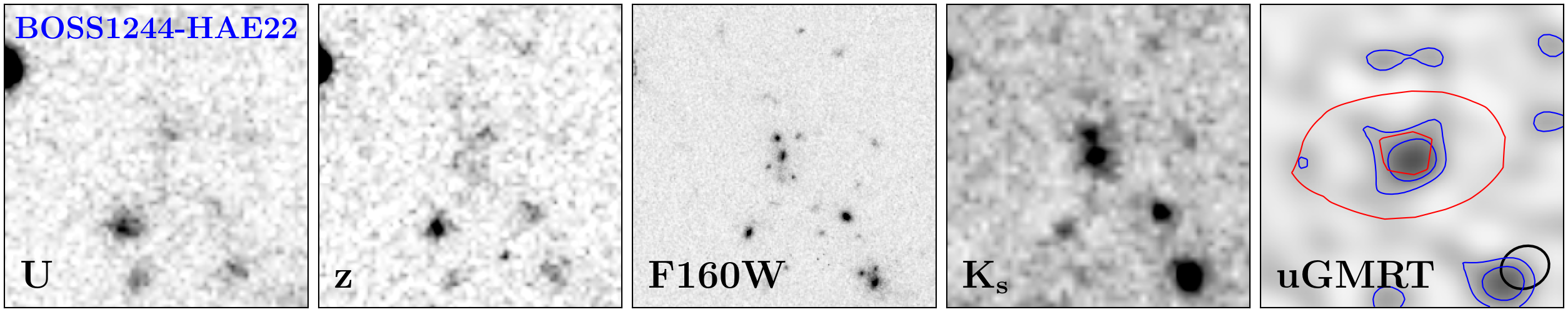}\par
        \includegraphics[width=\linewidth]{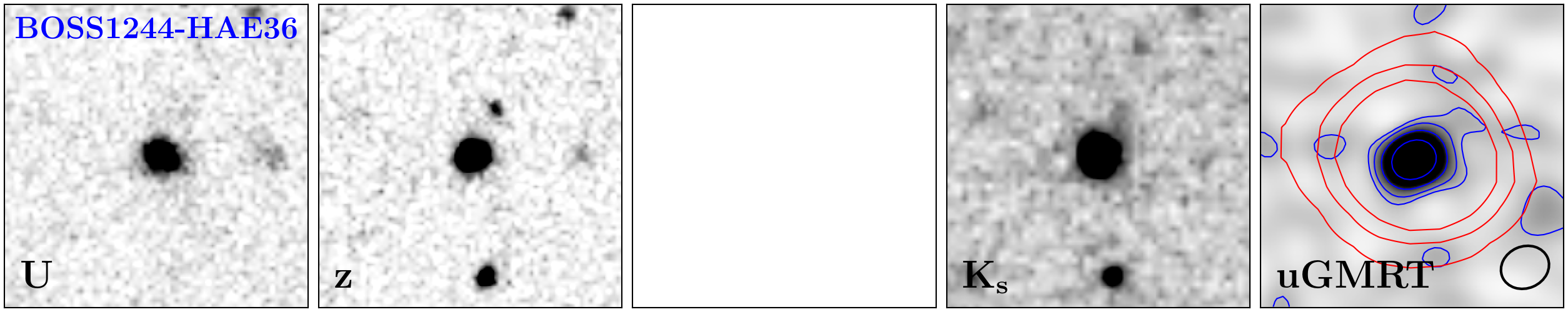}\par
        \includegraphics[width=\linewidth]{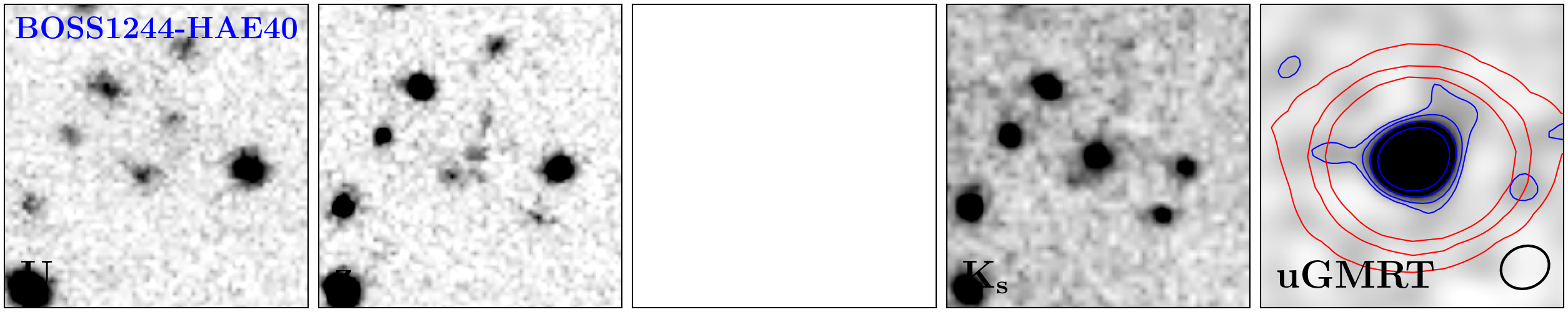}\par
        \includegraphics[width=\linewidth]{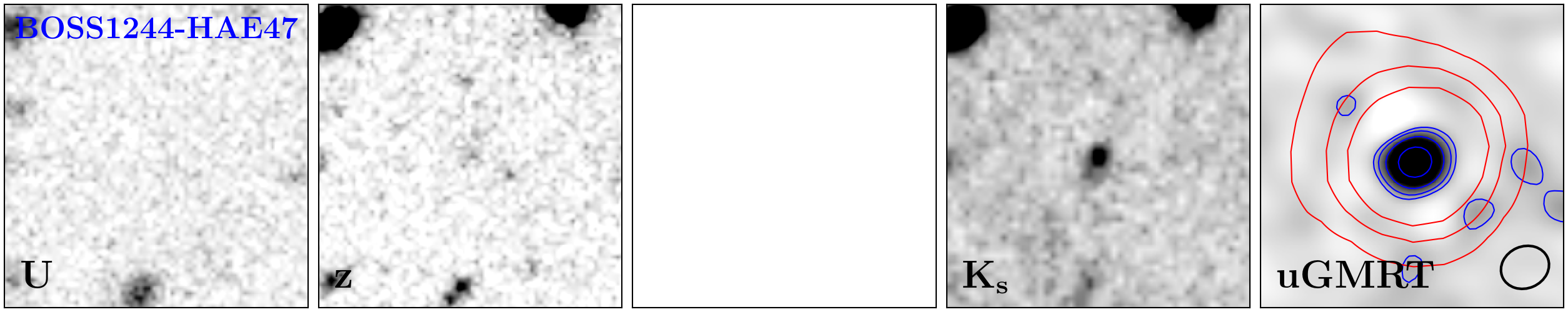}\par
        \includegraphics[width=\linewidth]{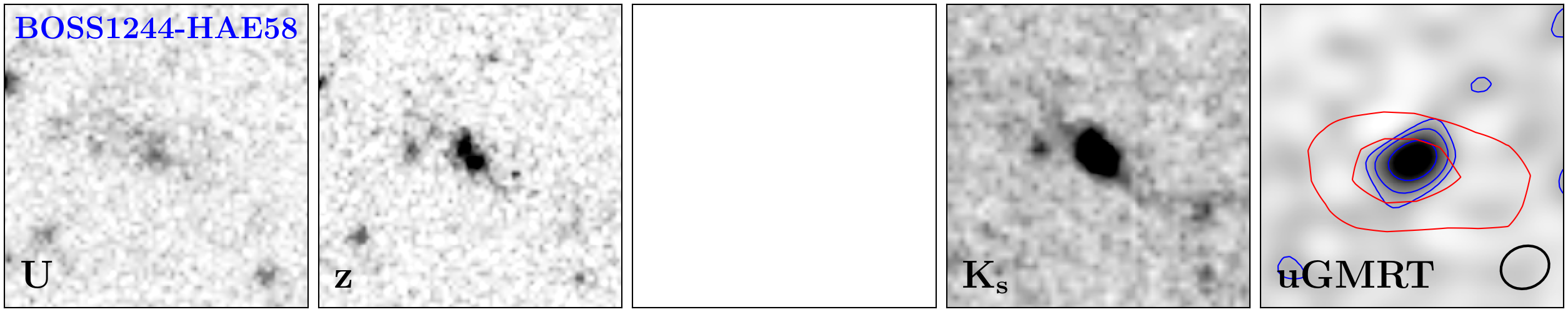}\par
        \includegraphics[width=\linewidth]{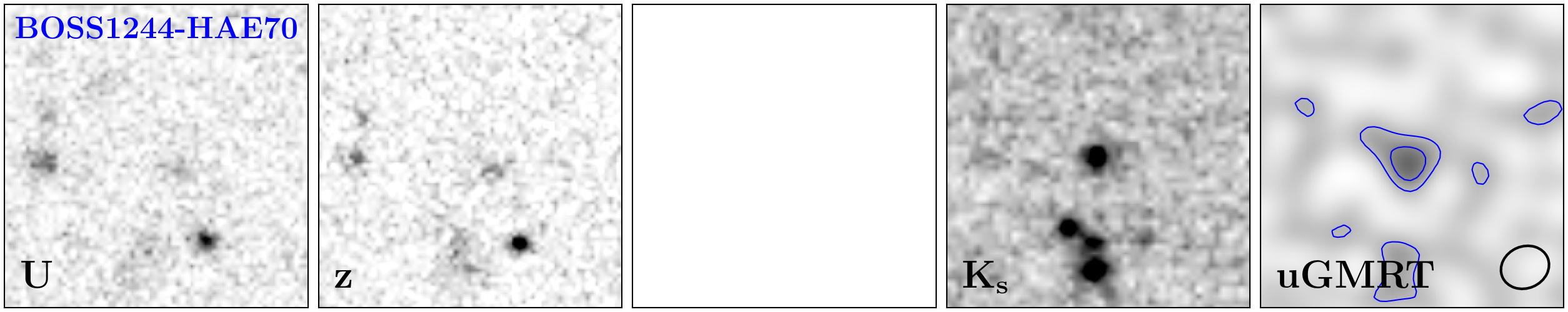}
    \end{minipage}
    \hfill
    \begin{minipage}[t]{0.495\textwidth}
        \includegraphics[width=\linewidth]{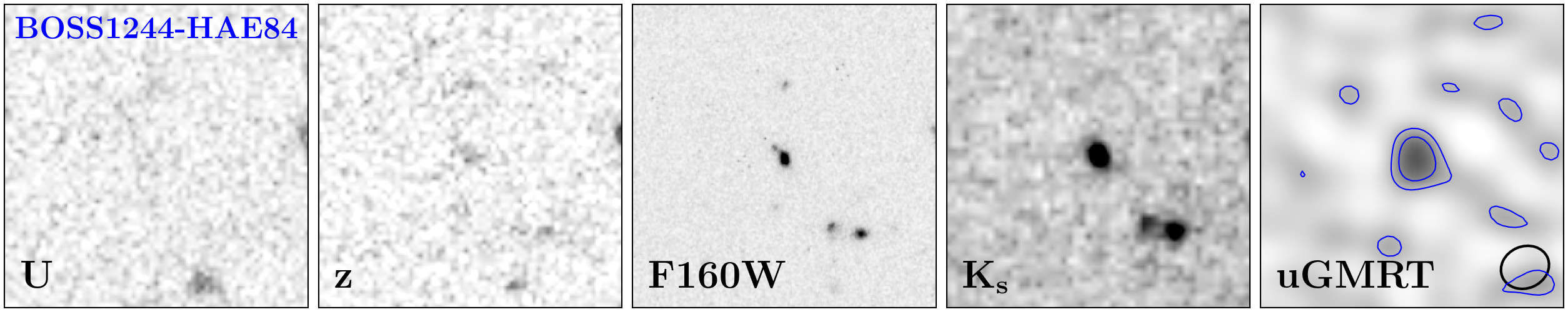}\par
        \includegraphics[width=\linewidth]{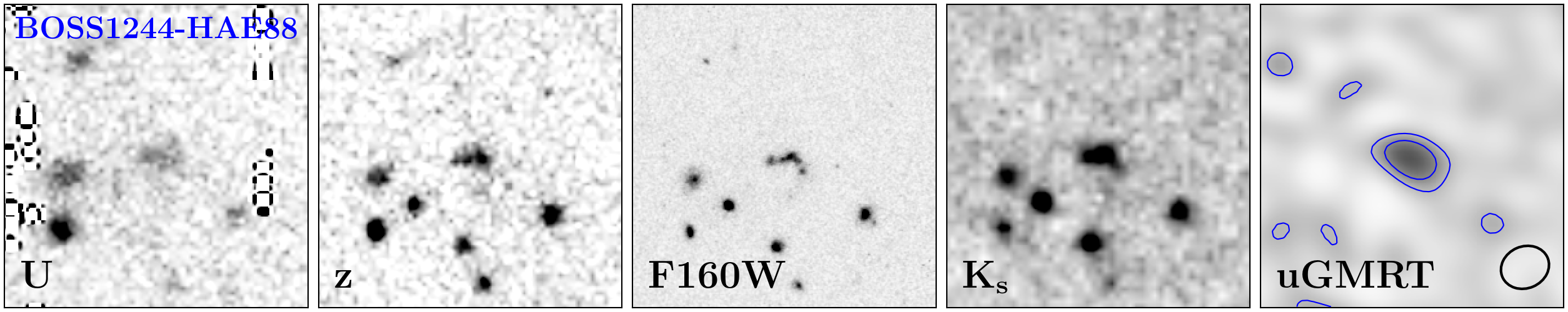}\par
        \includegraphics[width=\linewidth]{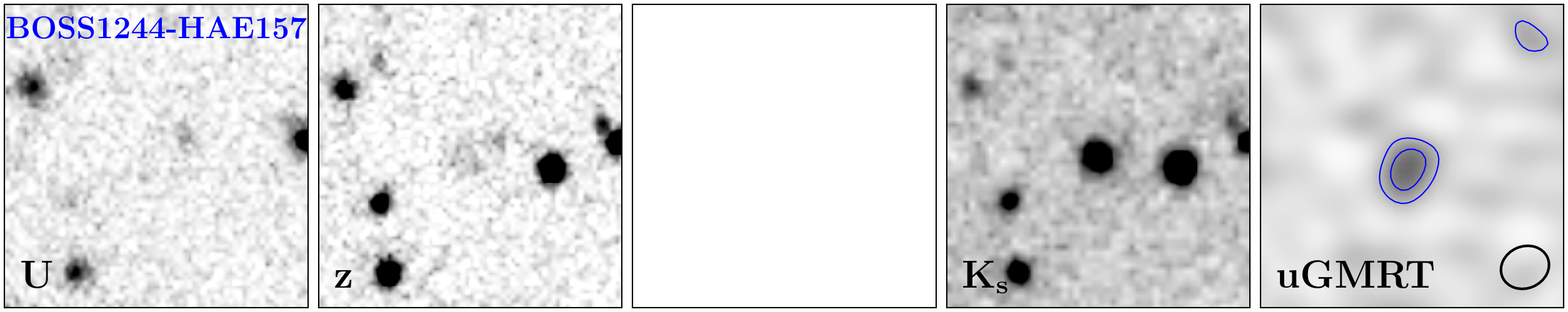}\par
        \includegraphics[width=\linewidth]{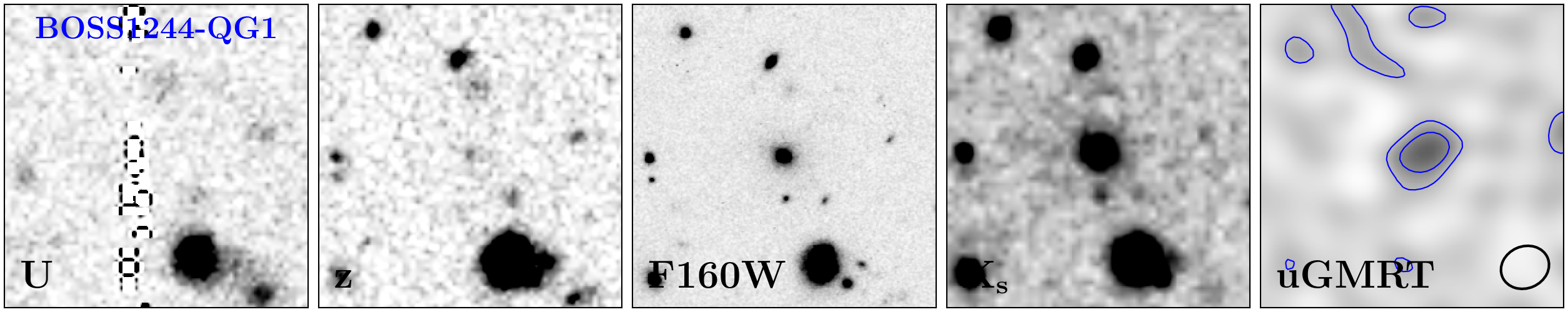}\par
        \includegraphics[width=\linewidth]{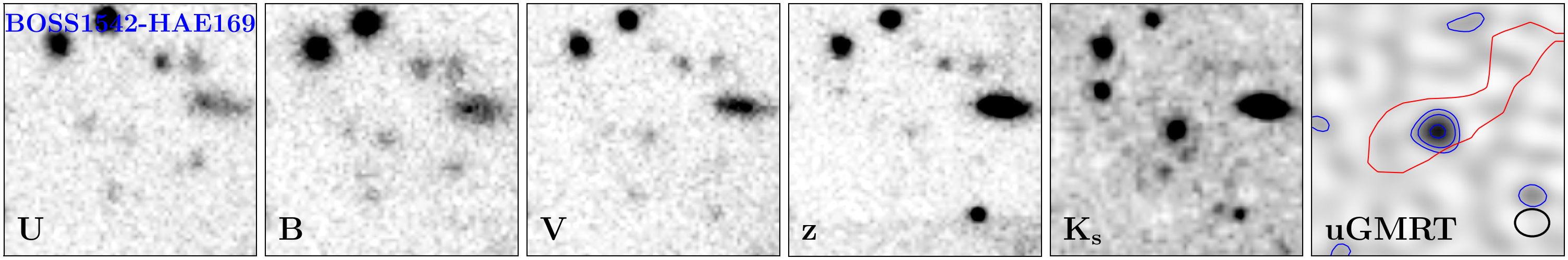}\par
        \includegraphics[width=\linewidth]{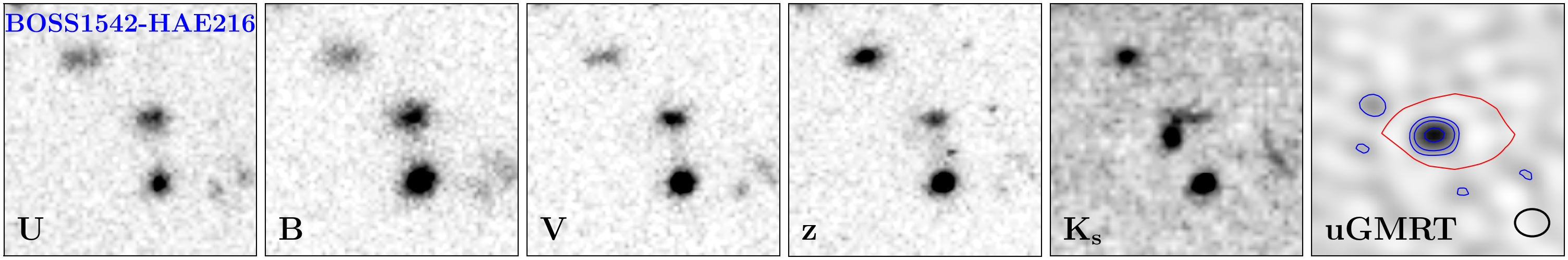}
    \end{minipage}
    \caption{ Multiwavelength cutouts of the twelve robust HAE radio counterparts and QG1. BOSS1244 rows show, from left to right, the $U$, $z$, F160W, $K_{\rm s}$, and uGMRT images; BOSS1542 rows show the $U$, $B$, $V$, $z$, $K_{\rm s}$, and uGMRT images. Blank F160W panels indicate regions without HST coverage. Each panel spans $15\arcsec\times15\arcsec$, with north up and east to the left. The rightmost panels show the uGMRT signal-to-noise maps, with blue contours at $[2,4,8,16]\sigma$ and the restoring beam indicated by the black ellipse. Red contours at $[3,6,9]\sigma$ show LoTSS emission when a LoTSS source lies within $3\arcsec$ of the target position.}
    \label{fig:gallery}
\end{figure*}

\subsection{HAE membership and physical properties}
\label{subsec:results_mass_sfr}

\citet{ZhengX2021} carried out deep CFHT/WIRCam $\rm H_2S(1)$ and $K_{\rm s}$ imaging of the BOSS1244 and BOSS1542 fields to search for HAEs at $z=2.24$. Using $\rm H_2S(1)<22.5$ mag, rest-frame ${\rm EW}_{\rm rest}>45$\,\AA, and a narrowband excess significance of $\Sigma>3$ as the selection criteria,  244 and 223 narrowband emitters were identified in BOSS1244 and BOSS1542, respectively. These parent samples can include both genuine HAEs at $z=2.24$ and interlopers with other emission lines falling within the $\rm H_2S(1)$ filter.  To identify the HAE members and remove potential interlopers, we further refined the samples using broadband colors \citep{ShiD2026}. For BOSS1542, we adopted the star-forming $Bz_{\rm s}K_{\rm s}$ criterion, $(z_{\rm s}-K_{\rm s})-(B-z_{\rm s})>-0.2$ \citep{Daddi2004}, extending the boundary to $>-0.8$ for very young stellar populations \citep{Lee2012}, together with the $U-B<0.8+0.2(B-z_{\rm s})$ criterion to reject possible $z>3$ interlopers \citep{Hatch2011}.  Among the 223 narrowband emitters, 188 are detected in all four $UBz_{\rm s}K_{\rm s}$ bands, of which 154 satisfy the color criteria. Since deep $B$-band imaging is unavailable for BOSS1244, the corresponding boundaries were transformed to the $Uz_{\rm s}K_{\rm s}$ plane.  Among the 244 narrowband emitters, 227 have the required three-band photometry, of which 203 satisfy the transformed criteria. We further incorporated available spectroscopic redshifts from ground-based near-infrared observations \citep{ShiD2021} and HST/WFC3 grism spectroscopy \citep{WangX2022}, adopting spectroscopic membership where available and the color-based classification otherwise.  Combining the spectroscopic and color-based classifications yields final samples of 210 and 160 HAE members in BOSS1244 and BOSS1542, respectively. Unless otherwise stated, the subsequent analyses focus on these HAE members.

Given the limited rest-frame optical and near-infrared photometric coverage, stellar masses were estimated from the $K_{\rm s}$ luminosity using a mass-to-light ratio that depends on $z_{\rm s}-K_{\rm s}$, following \citet{Koyama2013b}. The $z_{\rm s}-K_{\rm s}$ color brackets the redshifted 4000\,\AA{} break and provides an age-dependent correction to the mass-to-light ratio \citep{Kodama1999}. This estimator has been commonly adopted for HAEs in protocluster environments at similar redshifts \citep[e.g.,][]{Hayashi2012,Koyama2013b,Koyama2014} and yields stellar masses broadly consistent with those obtained from full spectral energy distribution (SED) fitting on average \citep{Koyama2013b}.

\begin{table*}[t]
    \caption{Physical and radio properties of the robust HAE counterparts and QG1.}
    \label{tab:radio-counterparts}
    \centering
    \setlength{\tabcolsep}{3.0pt}
    \begin{tabular}{lccccccc}
\toprule
Source & R.A. & Dec. & $z$ & $\log(M_\ast/M_\odot)$ & SFR & $S_{\rm peak}\pm\sigma_{\rm rms}$ & $S_{\rm total}\pm\sigma_{\rm total}$ \\
& (deg) & (deg) & & & ($M_\odot\,{\rm yr}^{-1}$) & ($\mu$Jy\,beam$^{-1}$) & ($\mu$Jy) \\
\midrule
BOSS1244-HAE19 & 190.91718 & 35.85264 & -- & 11.09 & 169.1 & $105.3\pm17.4$ & $124.0\pm34.7$ \\
BOSS1244-HAE22 & 190.91874 & 35.85430 & 2.246 & 10.99 & 134.4 & $111.0\pm16.5$ & $148.8\pm36.8$ \\
BOSS1244-HAE36 & 190.99545 & 35.87458 & -- & 11.67 & 1190.0 & $438.8\pm15.3$ & $443.4\pm26.7$ \\
BOSS1244-HAE40 & 191.00367 & 35.87274 & -- & 11.10 & 70.8 & $1067.2\pm14.6$ & $1062.9\pm25.1$ \\
BOSS1244-HAE47 & 191.05200 & 35.88663 & -- & 10.58 & 31.9 & $368.0\pm16.1$ & $345.3\pm26.3$ \\
BOSS1244-HAE58 & 190.91924 & 35.90234 & 2.217 & 11.37 & 190.4 & $171.3\pm12.9$ & $210.1\pm26.5$ \\
BOSS1244-HAE70 & 190.96409 & 35.91236 & -- & -- & -- & $66.1\pm12.0$ & $91.6\pm27.6$ \\
BOSS1244-HAE84 & 190.91533 & 35.92784 & 2.227 & 10.81 & 88.5 & $74.3\pm11.3$ & $85.1\pm22.1$ \\
BOSS1244-HAE88 & 190.87425 & 35.93384 & -- & 10.75 & 59.0 & $83.4\pm12.5$ & $103.4\pm25.8$ \\
BOSS1244-HAE157 & 191.10081 & 36.04387 & -- & 11.16 & 76.8 & $112.9\pm20.0$ & $155.1\pm45.4$ \\
BOSS1542-HAE169 & 235.79282 & 39.02643 & -- & 11.03 & 155.1 & $118.0\pm12.6$ & $128.8\pm23.7$ \\
BOSS1542-HAE216 & 235.72644 & 39.08526 & -- & -- & -- & $135.3\pm13.8$ & $143.7\pm25.1$ \\
BOSS1244-QG1 & 190.88350 & 35.91102 & 2.244 & 11.62 & 4.9 & $77.9\pm12.8$ & $116.9\pm31.2$ \\
\bottomrule
    \end{tabular}
    \tablefoot{The HAE coordinates, stellar masses, and extinction-corrected H$\alpha$-based SFRs were drawn from the HAE master catalog, with the masses and SFRs based on a Chabrier IMF. For HAE members without an independent spectroscopic redshift, the redshift entry is left blank and $z=2.24$ was adopted when deriving the listed physical quantities, including stellar masses and SFRs. QG1 properties are from \citet{ShiD2024}; its SFR was converted from the reported logarithmic value. Peak flux densities and local rms values were measured from the PyBDSF catalog, and total flux densities and their uncertainties are the PyBDSF integrated measurements. Dashes denote unavailable measurements.}
\end{table*}

We derived SFRs from the H$\alpha$ luminosities measured from the $\rm H_2S(1)$ and $K_{\rm s}$ photometry. Following \citet{ZhengX2021}, the H$\alpha$+[N\,II] line fluxes and rest-frame equivalent widths were obtained from the narrowband excess relative to the $K_{\rm s}$ continuum. The [N\,II]$\lambda\lambda6548,6583$ contribution was corrected using the empirical $[{\rm N\,II}]/{\rm H}\alpha$--${\rm EW}_{\rm rest}$ relation of \citet{Sobral2009,Sobral2012}, yielding mean $[{\rm N\,II}]/{\rm H}\alpha$ ratios of 0.16 and 0.15 for BOSS1244 and BOSS1542, respectively. 
Dust attenuation was corrected for each galaxy using the $A_{\rm H\alpha}$--$M_\ast$ relation \citep{Garn2010}, in which the attenuation depends on stellar mass. This relation has been shown to remain applicable to H$\alpha$-selected galaxies at $z\sim1.5$--2 \citep{Hayashi2011,Sawicki2012,Koyama2013b,Koyama2014}.
The resulting extinction-corrected H$\alpha$ luminosities were converted to SFRs following \citet{Kennicutt1998}. For consistency, both the stellar masses and SFRs were converted to a Chabrier IMF \citep{Chabrier2003}. With the available photometric coverage, stellar masses and H$\alpha$-based SFRs were estimated for 184 and 138 member galaxies in BOSS1244 and BOSS1542, respectively. A more detailed description of the HAE selection and derivation of these physical properties is presented in \citet{ShiD2026}.

The derived stellar masses and SFRs are subject to both statistical and systematic uncertainties.
In addition to photometric uncertainties, the principal systematics arise from the adopted mass-to-light ratio, the empirical [N\,II] correction, and the dust attenuation correction based on stellar mass. The latter has intrinsic scatter and may not fully capture variations among individual galaxies at high redshift.
Residual membership contamination also remains where spectroscopy is unavailable, and H$\alpha$ emission from AGN hosts may contain a non-star-forming contribution. These effects limit the accuracy of individual measurements, but our subsequent analysis relies primarily on ensemble trends and representative quantities such as median stellar masses and SFRs, making our conclusions less sensitive to uncertainties in individual measurements, although common systematic offsets may remain.

\subsection{Radio counterparts and AGN diagnostics}
\label{subsec:results_cpt}

Within the effective uGMRT areas, 211 and 148 of the parent narrowband emitters are covered in BOSS1244 and BOSS1542, respectively. Among these, 187 and 105 belong to the HAE member samples defined in Sect.~\ref{subsec:results_mass_sfr} based on the spectroscopic and color classifications described above. We searched for radio counterparts to these HAE members in the uGMRT images. 
We first cross-matched the HAE members with radio-source catalogs extracted blindly from the uGMRT images using PyBDSF \citep{Mohan2015}, independently of the HAE positions.
The catalogs were generated with island and peak thresholds of 3$\sigma$ and 5$\sigma$, respectively, and a matching radius of $1\arcsec$ was adopted. This yields ten radio counterparts in BOSS1244 and two in BOSS1542.

To search for fainter counterparts below the $5\sigma$ PyBDSF detection threshold, we performed forced photometry at the positions of all HAE members within the effective uGMRT areas. At each position, we scaled the normalized restoring beam to measure the peak flux density and performed a free Gaussian fit to estimate the integrated flux density. The uncertainties were estimated from the local rms for the beam-scaled measurements and from matched residual-injection trials for the Gaussian fits. We detect 14 and three HAE members at $\mathrm{S/N}\geq3$ in BOSS1244 and BOSS1542, respectively, including all $\geq5\sigma$ sources identified in the PyBDSF catalogs. For sources common to both methods, the Gaussian-fit peak and integrated flux densities differ from the corresponding PyBDSF measurements by only 2\% in the median. Point-source injection tests at 3, 4, and 5$\sigma$ recover the input flux densities with deviations below 3\%, confirming that the low-S/N measurements do not show a significant flux bias.

We therefore divided the radio counterparts into two classes. Sources detected in the PyBDSF catalogs are considered robust counterparts, for which we adopted the PyBDSF total flux densities. The additional sources detected only by forced photometry at $3\leq\mathrm{S/N}<5$ are considered potential counterparts, for which we adopted the Gaussian integrated flux densities when the fits are successful. Sources below 3$\sigma$ were treated as undetected and were included only in the subsequent ensemble and stacking analyses.

Finally, we applied the same radio measurements to the two massive QGs in BOSS1244 reported by \citet{ShiD2024}. QG1 has a robust radio counterpart and is recovered by forced photometry at $\mathrm{S/N}=6.3$, whereas QG2 remains undetected with $\mathrm{S/N}=1.2$. No counterpart is found by PyBDSF for any of the known quasars (QSOs) in either field \citep{ShiD2021}, and those within the usable uGMRT area all remain below 3$\sigma$ in the forced photometry.

The multiwavelength cutouts of the twelve robust HAE counterparts and the radio-detected QG1 are shown in Fig.~\ref{fig:gallery}, while the five potential HAE counterparts identified only through forced photometry are shown separately in Fig.~\ref{fig:potential-gallery} of Appendix~\ref{app:potential-counterparts}. The cutouts compare the available optical and near-infrared imaging with the uGMRT emission at each HAE or QG position. The coordinates, redshifts, stellar masses, SFRs, and uGMRT flux densities of the robust counterparts are summarized in Table~\ref{tab:radio-counterparts}. 

For comparison with the dusty galaxy population in the same fields, we also cross-matched the PyBDSF catalogs of $\geq5\sigma$ radio sources with the SMG samples using a radius of $7.5\arcsec$, corresponding to half the SCUBA-2 beam at 850\,$\mu$m \citep{ZhangY2022}. Within the effective areas, 14 of 42 SMGs in BOSS1244 and 21 of 54 in BOSS1542 have a radio counterpart at $\geq5\sigma$, corresponding to radio fractions of 33\% and 39\%, respectively. The combined fraction of 36\% is broadly consistent with the studies in blank fields \citep{An2018,An2019}. 
Four robust HAE counterparts are also associated with SMGs: BOSS1244-HAE36, BOSS1244-HAE70, BOSS1542-HAE169, and BOSS1542-HAE216. Among the potential counterparts, BOSS1244-HAE67 and BOSS1542-HAE133 are each associated with an SMG, with BOSS1244-HAE67 sharing the same SMG counterpart as BOSS1244-HAE70.
Thus, the six radio HAE counterparts are associated with five distinct SMGs, as BOSS1244-HAE67 and HAE70 are separated by only $5.7\arcsec$ and fall within the same SCUBA-2 beam, preventing the single-dish submillimeter emission from being uniquely assigned to either galaxy.

We quantified the radio excess using the criterion of \citet{Delvecchio2017},
$\log_{10}(L_{1.4\,\mathrm{GHz}}/{\rm SFR})>21.984(1+z)^{0.013}$, corresponding to a threshold of 22.323 at $z=2.24$. The integrated 1.26\,GHz flux densities were converted to rest-frame 1.4\,GHz luminosities assuming $S_\nu\propto\nu^\alpha$ with $\alpha=-0.7$ \citep{Delhaize2017}. Since direct infrared SFRs are unavailable for most sources, we used the H$\alpha$-based SFRs as a proxy and defined $\Delta_{\rm RE}\equiv\log_{10}(L_{1.4\,\mathrm{GHz}}/{\rm SFR}_{\mathrm{H}\alpha})-22.323$. BOSS1244-HAE40 and HAE47 lie more than 1 dex above the threshold and are identified as strong radio-excess AGN candidates. HAE22, HAE58, HAE84, HAE88, and HAE157 lie only 0.05--0.38 dex above the threshold and are regarded as marginal candidates, while HAE19, HAE36, and HAE169 fall below it; HAE70 and HAE216 cannot be classified because reliable H$\alpha$ SFRs are unavailable. QG1 also shows a strong radio excess when adopting its low SFR from \citet{ShiD2024}. Given that the \citet{Delvecchio2017} criterion was calibrated using infrared-based SFRs, we interpret these H$\alpha$-based classifications conservatively.

\begin{figure}[t]
    \centering
    \includegraphics[width=\linewidth]{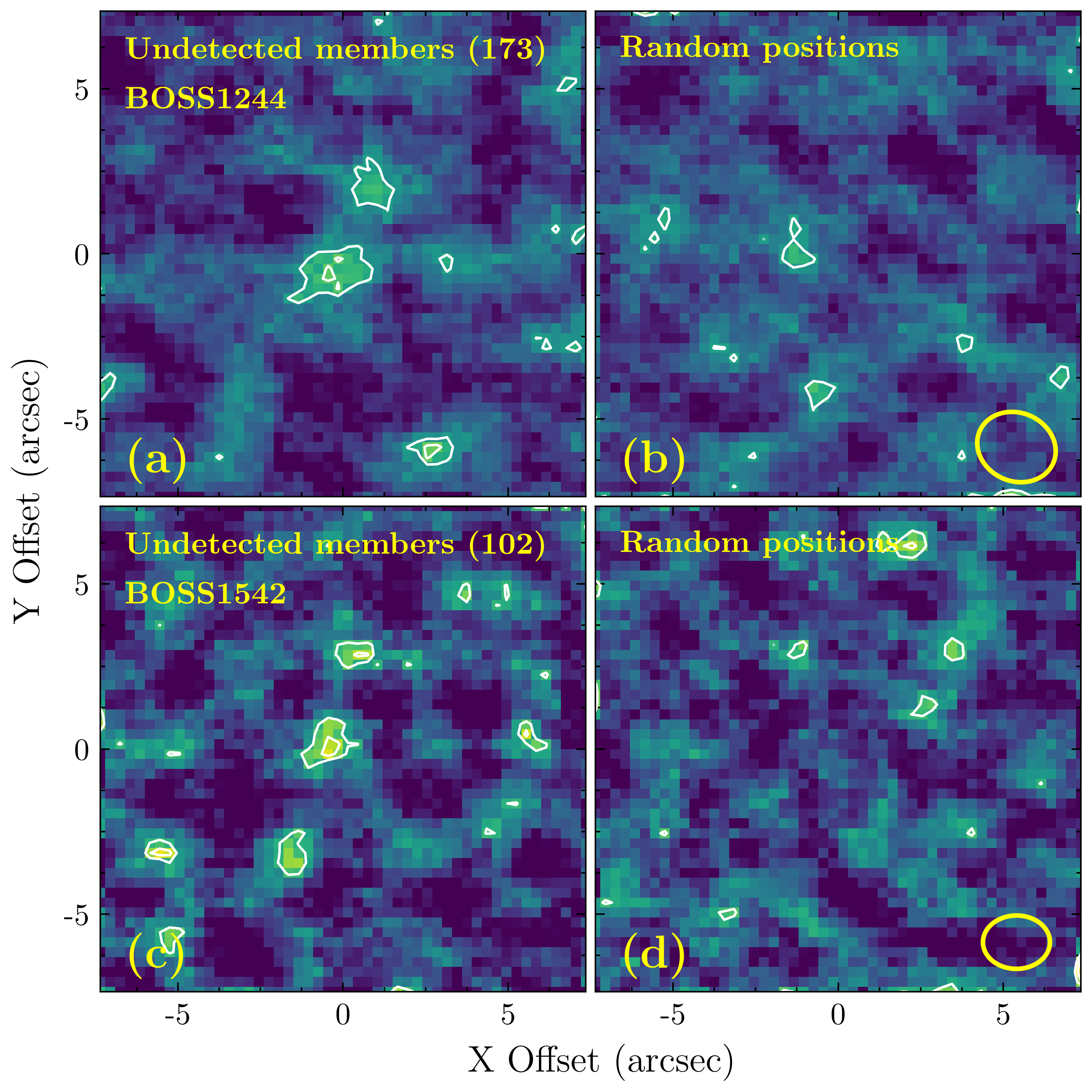}
    \caption{Median stacks of the primary-beam-corrected 1.26 GHz uGMRT radio continuum images for BOSS1244 (top) and BOSS1542 (bottom). Panels (a) and (c) show the radio-undetected HAE members, containing 173 and 102 sources, respectively. Panels (b) and (d) show representative matched random-position stacks using the same number of positions and matched primary-beam response and local rms distributions. Each panel spans $15\arcsec\times15\arcsec$, and all panels share the same color scale. White contours mark $[2,3,4]\sigma_{\rm img}$, where $\sigma_{\rm img}=1.5$ and $1.9~\mu{\rm Jy\,beam^{-1}}$ for BOSS1244 and BOSS1542, respectively. The yellow ellipses in the right-hand panels indicate the restoring beams.}
    \label{fig:full-stack-sequence}
\end{figure}

For the twelve robust HAE radio counterparts, we adopted the cataloged SCUBA-2 850\,$\mu$m flux densities for SMG detections and local $3\sigma$ upper limits for nondetections \citep{ZhangY2022}, and converted them to infrared SFRs using the templates of \citet{Magdis2012}. HAE40 and HAE47 remain above the radio-excess threshold, while the classifications of the other robust counterparts remain unchanged. These infrared-based estimates therefore provide additional support for the identification of HAE40 and HAE47 as strong radio-excess AGN candidates.

We then examined the radio spectral slopes by cross-matching the twelve robust HAE counterparts with the LoTSS-DR3 catalog \citep{Shimwell2026} using a $3\arcsec$ radius. Seven sources are detected at 144\,MHz, including five in BOSS1244 and both counterparts in BOSS1542. Their two-point spectral indices between 144\,MHz and 1.26\,GHz span $\alpha=-1.01$ to $-0.40$, with a median of $\alpha=-0.66$. BOSS1244-HAE40 has the flattest spectrum, with $\alpha=-0.40\pm0.03$, while BOSS1542-HAE169 has the steepest, with $\alpha=-1.01\pm0.22$. Given the coarser angular resolution of the LoTSS images, we used these spectral indices only as supporting evidence for the source classifications.

\begin{figure*}[t]
    \centering
    \includegraphics[width=0.49\textwidth]{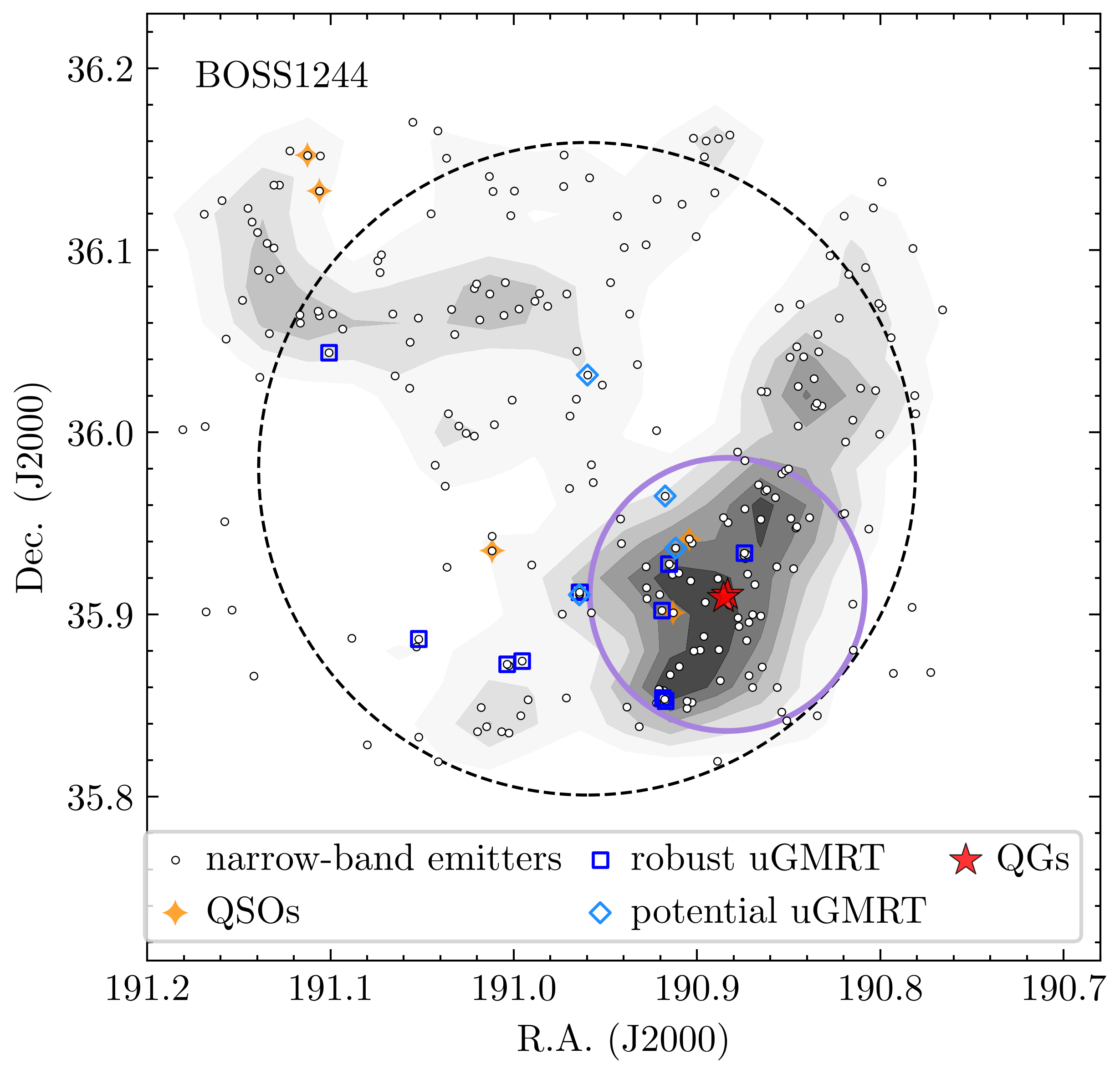}\hfill
    \includegraphics[width=0.49\textwidth]{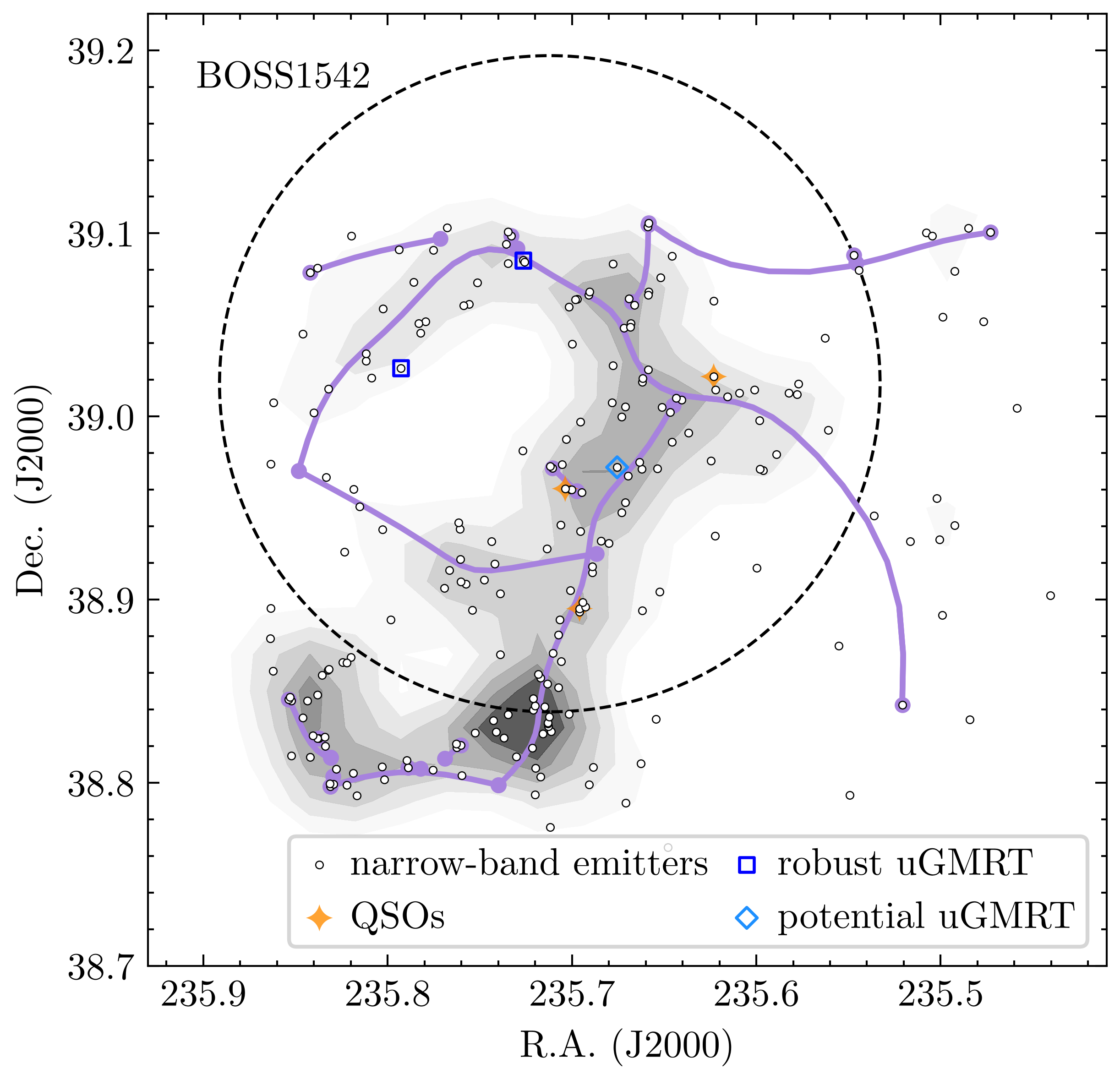}
    \caption{Spatial distributions and environmental measures used for the group stacking in BOSS1244 (left) and BOSS1542 (right). The gray filled contours trace the projected surface density of the parent narrowband emitters, and the open circles mark the individual sources. Blue squares and diamonds identify the robust ($\geq5\sigma$) and potential ($3\leq{\rm S/N}<5$) uGMRT counterparts, respectively; orange four-pointed stars mark QSOs, and the red stars in BOSS1244 mark the two massive QGs. The dashed circles indicate the effective uGMRT coverage. In BOSS1244, the solid purple circle marks the QG1-centered region used for environmental stacking, within which we use the projected distance from QG1 to characterize the environment. In BOSS1542, the purple curves show the filament skeleton identified with \texttt{DisPerSE}, and we use the minimum projected distance to the nearest filament segment.}
    \label{fig:environment-spatial}
\end{figure*}

Finally, the beam-averaged rest-frame brightness temperatures range from 33 to 529\,K, well below the maximum expected from star formation \citep{Condon1992,Morabito2022}, and therefore do not independently indicate AGN activity. Compact nuclear emission may nevertheless remain unresolved within the uGMRT beam. We therefore retain the radio-excess sources as AGN candidates, while the present data are insufficient for secure AGN identifications.

\subsection{Stacked emission of radio-undetected HAE members}
\label{subsec:results_stack}

To characterize the average faint radio emission below the individual detection threshold, we restricted the stacking analysis to HAE members without individual radio counterparts. Starting from the 187 and 105 HAE members within the effective uGMRT areas, we first excluded the ten and two robust $\geq5\sigma$ PyBDSF counterparts. We then removed an additional four and one potential counterparts identified through forced photometry, leaving 173 and 102 radio-undetected HAE members in BOSS1244 and BOSS1542, respectively. 

For each field, we extracted $15\arcsec\times15\arcsec$ radio cutouts centered on the HAE member positions and combined them using the median. Figure~\ref{fig:full-stack-sequence} shows the resulting median stacks for these two samples.
We measured the stacked peak flux densities using the same beam-template photometry described in Sect.~\ref{subsec:results_cpt}, while accounting for the measured radio--$K_{\rm s}$ astrometric scatter. As a consistency check, we also measured restoring-beam aperture fluxes with the corresponding aperture correction; the two measurements agree within $1\sigma$ for all stacks. Uncertainties were estimated from 500 bootstrap resamplings. To evaluate the detection significance, we additionally generated 500 matched random-position stacks with the same number of positions and matched primary-beam response and local rms distributions as the corresponding HAE sample. The significance of the observed stack was evaluated relative to the distribution of peak flux densities recovered from these random stacks.

The radio-undetected members yield stacked peak flux densities of $4.17^{+1.36}_{-1.31}\,\mu$Jy in BOSS1244 and $4.05^{+1.56}_{-1.72}\,\mu$Jy in BOSS1542. Relative to the matched random-position stacks, these signals are detected at $2.8\sigma$ and $3.2\sigma$, respectively. Both stacks are consistent with unresolved emission, and we therefore adopted the peak flux densities as our fiducial measurements. We converted the stacked flux densities to rest-frame 1.4\,GHz luminosities assuming $S_\nu\propto\nu^\alpha$ with $\alpha=-0.7$. Using the radio--SFR calibration of \citet{Novak2017}, the resulting luminosities are comparable to those expected from the median H$\alpha$-based SFRs of the same samples. Star formation can therefore account for the average faint radio emission at the present sensitivity, although a contribution from weak AGN activity cannot be excluded.

\subsection{Environmental dependence of the stacked radio emission}
\label{subsec:results_group_stack}

The distinct internal structures of BOSS1244 and BOSS1542 provide an opportunity to examine whether the faint radio emission of their HAE populations varies with environment \citep{ZhengX2021,ShiD2021,LiuS2023,ShiD2024}. For the environmental analysis, we define radio-faint HAE members as those remaining after excluding the robust $\geq5\sigma$ PyBDSF counterparts. We characterized the environment using the projected distance from QG1 in BOSS1244 and the minimum projected distance to the nearest filament segment in BOSS1542, as illustrated in Fig.~\ref{fig:environment-spatial}. In BOSS1244, QG1 lies at the center of the prominent HAE overdensity \citep{ShiD2024}. We considered the largest circular region centered on QG1 that remains fully covered by the effective uGMRT area, corresponding to a radius of $5\arcmin$. Within this region, 77 radio-faint HAE members were divided by increasing distance from QG1 into three nearly equal groups of 26, 26, and 25 sources. The three bins span $0.13$--$2.05\arcmin$, $2.06$--$3.62\arcmin$, and $3.64$--$4.94\arcmin$, with median distances of $1.40\arcmin$, $2.85\arcmin$, and $4.10\arcmin$, respectively. BOSS1542 instead exhibits a predominantly filamentary structure. We constructed the filament skeleton with \texttt{DisPerSE} \citep{Sousbie2011a} using the two-dimensional projected distribution of the parent narrowband emitters, and characterized the environment by the minimum projected distance, $d_{\rm fil}$, to the nearest filament segment. Following the redshift evolution of filament radii found in simulations \citep{WangW2024}, we restricted the analysis to $d_{\rm fil}\leq1\arcmin$, corresponding to approximately $1.5$ comoving Mpc at $z=2.24$. The resulting 85 radio-faint members were divided into two groups of 43 and 42 sources spanning $0.00$--$0.26\arcmin$ and $0.27$--$0.98\arcmin$, with median distances of $0.13\arcmin$ and $0.51\arcmin$, respectively.

\begin{figure*}[t]
    \sidecaption
    \hbox to 12cm{\includegraphics[width=5.9cm]{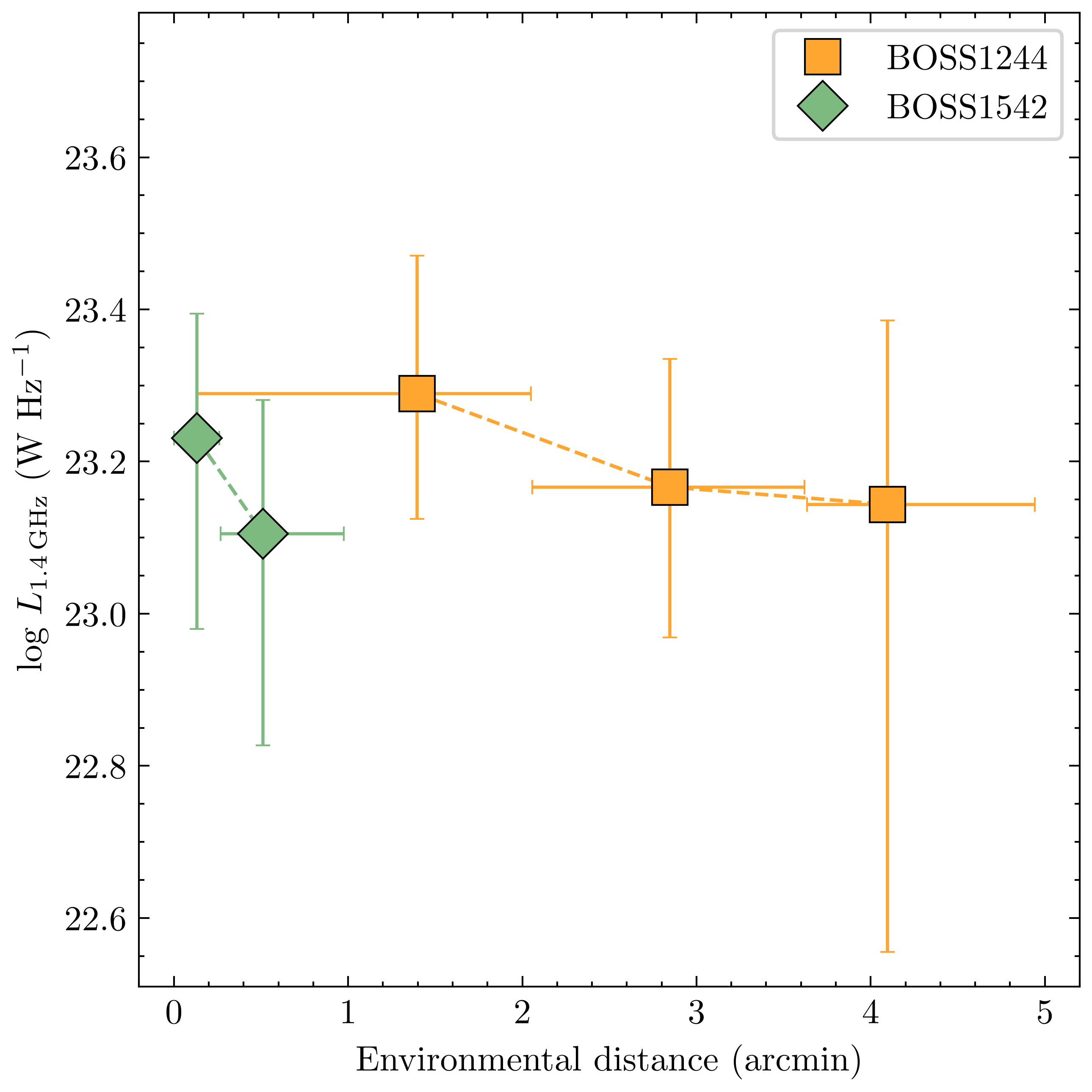}\hfil\includegraphics[width=5.9cm]{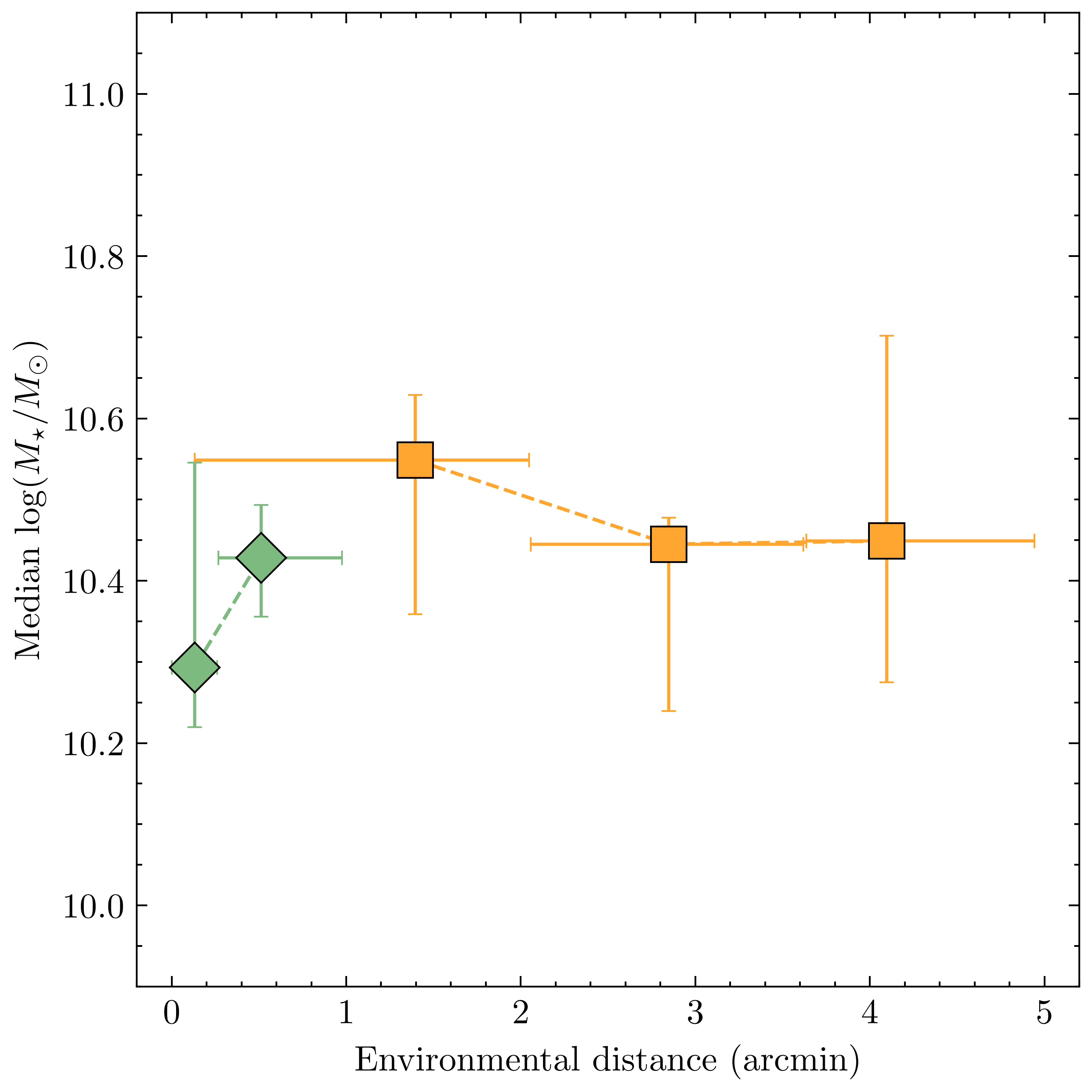}}
    \caption{Environmental group properties of the radio-faint HAE members. Left: rest-frame 1.4\,GHz luminosity inferred from the median-stacked peak flux density as a function of the projected distance from QG1 in BOSS1244 (orange squares) and from the nearest filament spine in BOSS1542 (green diamonds). The vertical bars show the 16th--84th percentiles from source bootstrap resampling, and the horizontal bars span the full distance range of each fixed bin. Right: median stellar mass for the same groups, using only sources with valid measurements. The vertical bars are the 16th--84th percentiles of the median from 10,000 bootstrap realizations; they are not the 16th--84th percentile scatter of the individual galaxies. Dashed lines connect the group measurements only and do not represent fitted relations.}
    \label{fig:environment-stacks-properties}
\end{figure*}

We applied the same median stacking and peak flux measurements as described in Sect.~\ref{subsec:results_stack}. The three groups in BOSS1244 yield stacked flux densities of $7.83^{+4.05}_{-2.47}$, $5.90^{+2.79}_{-2.16}$, and $5.60^{+4.17}_{-4.15}\,\mu$Jy, while the near- and far-filament groups in BOSS1542 give $6.84^{+3.13}_{-3.00}$ and $5.12^{+2.56}_{-2.42}\,\mu$Jy, respectively. We converted these measurements to rest-frame 1.4\,GHz luminosities assuming $\alpha=-0.7$; the resulting luminosities are shown in the left panel of Fig.~\ref{fig:environment-stacks-properties}. In both fields, the largest stacked radio luminosity is found in the densest environment, namely the innermost bin in BOSS1244 and the near-filament bin in BOSS1542. 
The corresponding flux differences between the innermost and outermost bins are $2.23~\mu$Jy in BOSS1244 and $1.72~\mu$Jy in BOSS1542. Bootstrap resampling shows that both differences remain consistent with zero. To assess whether the observed environmental differences could arise by chance, we performed permutation tests by randomly shuffling the association between galaxies and environmental distances while preserving the original group sizes. The tests give two-sided probabilities of $p=0.60$ and $p=0.66$, respectively. A linear fit to the three BOSS1244 bins gives a slope of $-0.84~\mu{\rm Jy\,arcmin^{-1}}$, with a permutation probability of $p=0.59$.
Injection tests using sources of equal flux recover all groups with biases below $0.5\,\mu$Jy and show no evidence that the stacking procedure introduces a systematic environmental trend.

We further compared the radio measurements with the stellar masses of the same groups. Valid measurements are available for 23, 23, and 21 galaxies in the three BOSS1244 bins and for 35 and 36 galaxies in the two BOSS1542 bins. The median $\log(M_\ast/M_\odot)$ values are [10.55, 10.45, 10.45] in BOSS1244 and [10.29, 10.43] in BOSS1542, with their bootstrap uncertainties shown in the right panel of Fig.~\ref{fig:environment-stacks-properties}. We used the Spearman rank correlation test to assess whether stellar mass shows a monotonic dependence on environmental distance. The Spearman coefficients are $\rho=0.10$ ($p=0.43$) in BOSS1244 and $\rho=0.00$ ($p=0.98$) in BOSS1542, indicating no significant correlation in either field.

To assess whether the absolute levels of the stacked radio emission are broadly consistent with star formation, we converted the median H$\alpha$-based SFR of each group to the expected rest-frame 1.4 GHz luminosity using the radio--SFR calibration \citep{Novak2017}. The resulting expected $\log(L_{\rm 1.4\,GHz}/{\rm W\,Hz^{-1}})$ values are [23.35, 23.19, 23.27] for the three BOSS1244 groups and [23.22, 23.23] for the two BOSS1542 groups. For comparison, the corresponding luminosities measured from the radio stacks are [23.28, 23.15, 23.13] and [23.22, 23.09], respectively. The expected luminosities are therefore comparable to the stacked measurements, indicating that their absolute levels do not require an additional radio component beyond that expected from star formation. This comparison remains approximate because the H$\alpha$-based SFRs are subject to uncertainties in dust attenuation and possible AGN contamination, with the latter potentially overestimating the true star-forming contribution. Given these uncertainties, the observed radio luminosities are consistent with being dominated by star formation at the present sensitivity, while a weak AGN contribution cannot be excluded. Overall, the current data provide only tentative evidence for an environmental modulation of faint radio emission, which requires deeper observations to establish.

\section{Discussion}
\label{sec:discuss}

\subsection{Nature of the radio emission in protocluster members}
\label{subsec:discuss-radio}

The individual radio detections suggest that the radio emission of protocluster HAEs has a heterogeneous origin. The strongest evidence for an AGN contribution comes from BOSS1244-HAE40 and HAE47, together with the radio-detected QG1, all of which show strong radio excesses relative to their H$\alpha$-based SFRs.
For HAE40 and HAE47, the classifications remain unchanged when SFRs inferred from the SCUBA-2 measurements are adopted, strengthening the case that their radio emission cannot be explained by star formation alone. 
If either the H$\alpha$ or single-band submillimeter emission contains a nonnegligible AGN contribution, the corresponding SFR would overestimate the true star-forming contribution and thus provide a conservative upper limit. The intrinsic radio excesses of HAE40 and HAE47 would therefore be even stronger after accounting for such contamination, further supporting an AGN origin.
H$\alpha$ contamination by low-luminosity AGNs is indeed demonstrated for massive HAEs in the Spiderweb protocluster \citep{Shimakawa2024a}.
The more marginal radio-excess sources are less secure. In particular, the criterion of \citet{Delvecchio2017} was calibrated using SFRs derived from infrared emission, whereas our primary estimates are based on H$\alpha$. Obscured star formation could cause the H$\alpha$-based SFR to be underestimated and hence artificially enhance the apparent radio excess. The marginal excesses should therefore be interpreted as possible AGN contributions rather than firm classifications.

The available radio spectral information does not remove this ambiguity. The spectral indices between 144\,MHz and 1.26\,GHz span $\alpha=-1.01$ to $-0.40$, with substantial overlap between the values commonly observed for star-forming galaxies and radio AGNs \citep{CalistroRivera2017}. BOSS1244-HAE40 is an exception in the sense that its relatively flat spectrum, $\alpha=-0.40\pm0.03$, provides additional support for a compact radio component. However, differences in angular resolution between LoTSS and uGMRT limit the diagnostic power of these spectral indices for the remaining sources. The brightness temperatures measured within the uGMRT beam provide a similarly one-sided constraint: their low values do not require an AGN, but compact nuclear emission can remain diluted within the synthesized beam \citep{Condon1992,Morabito2022}. Higher resolution radio imaging is therefore needed to determine whether such compact components are present in the marginal candidates.

The stacked emission probes a different regime from these individually detected sources. Positive signals remain after removing both the robust and potential counterparts, indicating a positive average radio signal from the radio-undetected HAE member population. At the same time, the stacked radio luminosities are comparable to those expected from the median H$\alpha$-based SFRs using the radio--SFR calibration \citep{Novak2017}. Star formation is therefore sufficient to explain the average radio emission at the present sensitivity. This does not require the faint HAE population to be free of nuclear activity: a weak AGN contribution present in only a subset of galaxies could have little effect on the median stack and remain hidden within the intrinsic scatter of the radio emission associated with star formation.

Taken together, the observations are consistent with a picture in which the bright end of the radio HAE population contains a small number of clear radio-excess AGN candidates, whereas the much fainter population is on average consistent with radio emission associated with star formation. Weak nuclear activity may still be present among the latter, but the current data cannot determine its incidence or contribution. This distinction is important because the stacked signal need not have the same physical origin as the strongest individually detected sources. Radio observations at multiple frequencies and higher angular resolution, combined with SFR estimates from infrared data or full SED fitting, are required to isolate compact AGN emission and separate it from star formation more directly.

\subsection{Possible environmental modulation of faint radio emission}
\label{subsec:discuss-environment}

The environmental stacking analysis shows the same qualitative ordering in the two protoclusters despite their different large-scale structures. In BOSS1244, the largest stacked radio luminosity is found in the group closest to QG1 and the central HAE overdensity, whereas in BOSS1542 it is found closer to the filament spine. Because all individually detected $\geq5\sigma$ counterparts have been removed from these samples, this ordering is not driven by a small number of bright radio sources. Injection tests using sources of equal flux also show no evidence that the stacking procedure introduces a systematic trend. Nevertheless, the statistical evidence remains weak: the differences between the innermost and outermost groups are consistent with zero, with permutation probabilities of $p=0.60$ and $p=0.66$ for BOSS1244 and BOSS1542, respectively, while the fitted radial slope in BOSS1244 is likewise insignificant. The current observations therefore do not establish a significant dependence of faint radio emission on environment.

The stellar masses provide further context for this tentative ordering. They show no significant monotonic dependence on environmental distance in either protocluster, suggesting that the radio ordering is not simply driven by a corresponding trend in stellar mass. The stacked radio luminosities are also broadly consistent with those expected from the median H$\alpha$-based SFRs in all five groups. However, as discussed above, these SFR estimates remain uncertain because of dust attenuation and possible AGN contamination. In particular, an AGN contribution to the H$\alpha$ emission would overestimate the true star-forming contribution.  The present data therefore cannot distinguish between normal variations in radio emission associated with star formation and an additional weak AGN component whose incidence or strength may vary with environment.

If this qualitative ordering is confirmed by deeper observations, enhanced weak AGN activity would provide one plausible interpretation. Dense protocluster environments can increase the frequency of interactions between galaxies and mergers, which may promote gas inflow toward galactic nuclei \citep{HopkinsP2006,Angles-Alcazar2017}. Such processes are particularly relevant to BOSS1244 and BOSS1542, where the merger rates of massive HAEs are already enhanced relative to the field \citep{LiuS2023}. The incidence or duty cycle of weak AGN activity could therefore increase toward the densest regions without producing individually detectable radio excesses in most galaxies. Alternatively, environmental changes in obscured star formation, magnetic field strength, or cosmic ray confinement could modify the radio emission associated with star formation. The present data do not allow these possibilities to be distinguished.

The contrasting structures of the two protoclusters provide additional context for this tentative behavior. BOSS1244 contains two massive QGs, including QG1 at the central overdensity \citep{ShiD2024}, and an excess of early-type star-forming galaxies is reported around this region \citep{Golden-Marx2026}. These observations have been interpreted as possible evidence for an emerging cluster core. BOSS1542, in contrast, has a more extended filamentary structure and a higher merger rate than BOSS1244 \citep{LiuS2023}. Distance from QG1 in BOSS1244 and distance from the filament spine in BOSS1542 therefore trace physically different aspects of the assembling large-scale structure. The similar radio ordering found in these two distinct environments is suggestive, but does not imply that the same physical mechanism is operating in both systems.

Several limitations prevent a stronger conclusion. The environmental bins contain only a few tens of galaxies, resulting in large bootstrap uncertainties in the stacked radio measurements. 
In addition, the intrinsic scatter in the relation between infrared and radio emission is larger than the residual differences considered here.
In BOSS1542, the filament distance is also a projected quantity, so galaxies separated along the line of sight may be assigned similar environments. The present analysis therefore provides a tentative indication, rather than a detection, of environmental modulation in faint radio emission. Deeper radio imaging at multiple frequencies and with sufficient angular resolution to isolate compact nuclear components, together with SFR estimates from infrared data or full SED fitting, is needed to determine whether the observed ordering reflects enhanced weak AGN activity or variations in radio emission associated with star formation.

\section{Summary}
\label{sec:summary}

We presented deep 1.26\,GHz uGMRT observations of the two $z=2.24$ protoclusters, BOSS1244 and BOSS1542, and investigated the radio properties of their H$\alpha$-selected member galaxies. Our main results are summarized as follows:

\begin{enumerate}

\item 
Among the 187 and 105 HAE members covered by the effective uGMRT areas in BOSS1244 and BOSS1542, respectively, we identify ten and two robust radio counterparts at $\geq5\sigma$, together with four and one additional potential counterparts detected at $3\leq{\rm S/N}<5$ through forced photometry. BOSS1244-HAE40, HAE47, and the massive QG1 show strong radio excesses relative to their H$\alpha$-based SFRs and are identified as strong radio-excess AGN candidates. Seven of the twelve robust HAE counterparts are also detected in LoTSS-DR3, with spectral indices spanning $\alpha=-1.01$ to $-0.40$. The relatively flat spectrum of HAE40 provides additional support for an AGN contribution, while the spectral indices of the full sample do not provide a clean separation between AGN and radio emission associated with star formation.

\item 
Median stacking reveals faint radio emission from radio-undetected HAE members.
After removing both the robust and potential radio counterparts, the remaining 173 and 102 radio-undetected members yield positive stacked signals detected at $2.8\sigma$ in BOSS1244 and $3.2\sigma$ in BOSS1542, respectively.
Their stacked radio luminosities are comparable to those expected from the median H$\alpha$-based SFRs using the radio--SFR calibration \citep{Novak2017}. The average faint radio emission can therefore be explained by star formation at the present sensitivity, although a contribution from weak AGN activity cannot be excluded.

\item 
We further examined whether the faint radio emission varies within the two protoclusters. The largest stacked radio luminosity is found closest to QG1 and the central HAE overdensity in BOSS1244, and closer to the filament spine in BOSS1542. However, the differences between the environmental groups are not statistically significant, with permutation probabilities of $p=0.60$ and $p=0.66$, respectively, while the radial slope in BOSS1244 is also insignificant. Stellar mass shows no significant monotonic dependence on environmental distance. The current data therefore provide tentative evidence that faint radio emission is enhanced toward the denser regions.

\end{enumerate}

Taken together, our results identify a small population of strong radio-excess AGN candidates among individually detected protocluster members, while the much fainter HAE population is consistent on average with radio emission associated with star formation. The similar environmental ordering seen in both systems is suggestive, but its physical origin cannot yet be established. Radio observations at multiple frequencies and higher angular resolution, together with SFR estimates from infrared data or full SED fitting, are needed to determine whether the apparent enhancement toward dense cores and filaments reflects increased weak AGN activity, variations in radio emission associated with star formation, or a combination of both.

\begin{acknowledgements}
This work is supported by the National Science Foundation of China (grant no. 12603019) and the Postdoctoral Fellowship Program of CPSF under Grant Number GZC20261726.
This work is supported by the National Key Research and Development Program of China (2023YFA1608100), the National Science Foundation of China (12233005, 12173088), the China Manned Space Program with grants nos. CMS-CSST-2025-A08 and CMS-CSST-2025-A20, in part by Office of Science and Technology, Shanghai Municipal Government (grant Nos. 24DX1400100, ZJ2023-ZD-001).
This work was supported by National Natural Science Foundation of China (grant No. 12525302 and 12141301), National Key R\&D Program of China (grant No. 2023YFA1605600), Basic Research Program of Jiangsu (grant No. BK20250001), the Fundamental Research Funds for the Central Universities with grant No. KG202502, and the China Manned Space Program with grant No. CMS-CSST-2025-A04. T.W. acknowledges the support from the New Cornerstone Science Foundation through the XPLORER PRIZE.
FXA acknowledges the support from the National Natural Science Foundation of China (12303016) and the Natural Science Foundation of Jiangsu Province (BK20242115).
Q.T. acknowledges support from the NSFC (grants 12573013, 12033004).
This work is supported by the start-up funding of the Anhui University of Science and Technology (2024yjrc104), the National Science Foundation of China (grant no. 12303015), the Natural Science Foundation of Jiangsu Province (BK20231106) and the China Manned Space Program with grant No. CMS-CSST-2025-A20.
We thank the staff of the uGMRT that made these observations possible. The uGMRT is run by the National Centre for Radio Astrophysics of the Tata Institute of Fundamental Research.

\end{acknowledgements}

\bibliographystyle{aa}
\bibliography{aa}

\begin{appendix}
\nolinenumbers

\onecolumn
\section{Potential HAE radio counterparts}
\label{app:potential-counterparts}

\begin{figure}[ht!]
    \centering
    \includegraphics[width=0.55\textwidth]{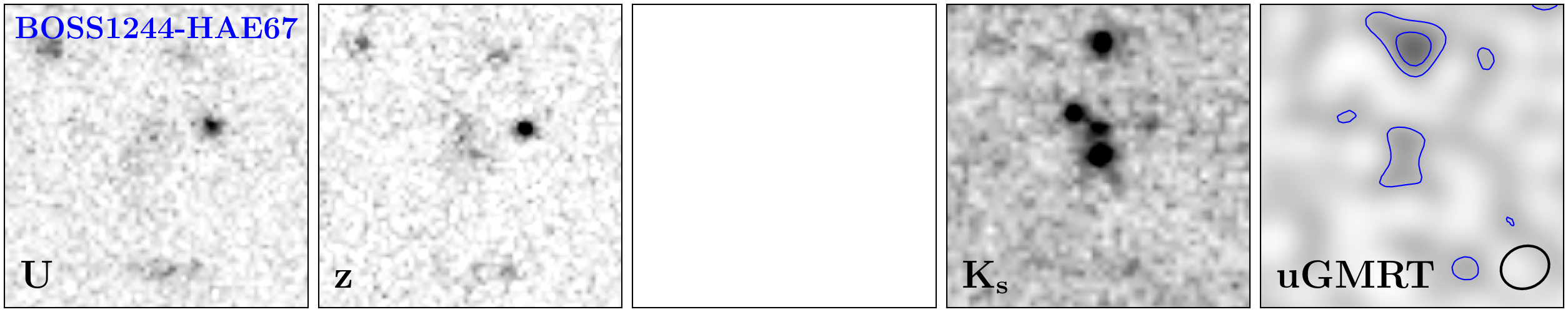}\par
    \includegraphics[width=0.55\textwidth]{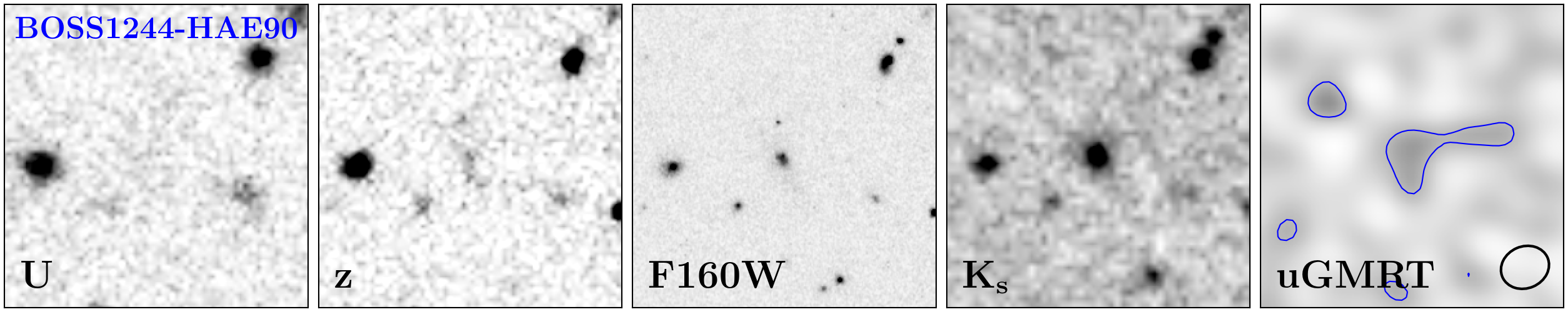}\par
    \includegraphics[width=0.55\textwidth]{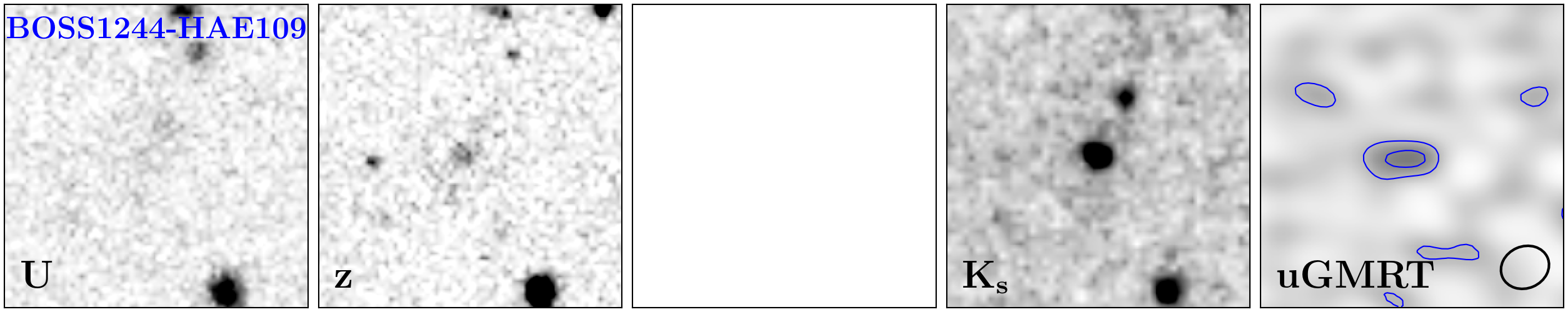}\par
    \includegraphics[width=0.55\textwidth]{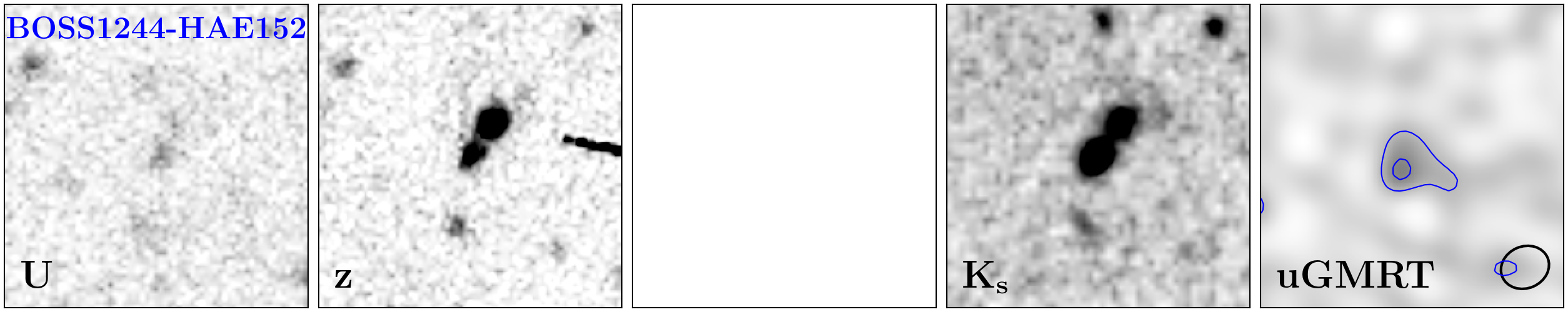}\par
    \includegraphics[width=0.55\textwidth]{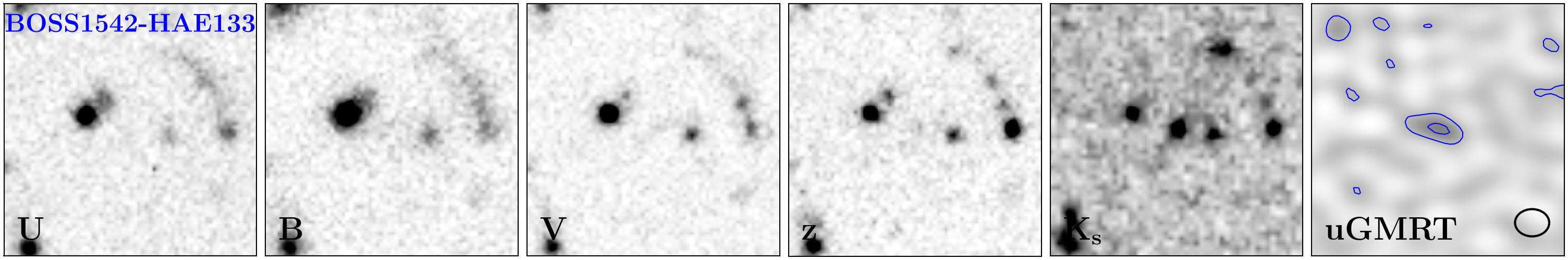}
    \caption{
    Multiwavelength cutouts of the five potential HAE radio counterparts selected by forced photometry at $3\leq{\rm S/N}<5$. The bands, image orientation, uGMRT and LoTSS contours, and panel size are the same as in Fig.~\ref{fig:gallery}.}
    \label{fig:potential-gallery}
\end{figure}

\end{appendix}
\end{document}